\documentclass[aip,reprint]{revtex4-1}

\usepackage{graphicx}

\usepackage{bm}% bold math
\begin{document}

%\preprint{}

\title{THE STEADY INCOMPRESSIBLE IDEAL FREE-BOUNDARY FLOWS OF A  HYDROMAGNETIC STAR}%Title of paper



\author{B. C. Low}
%\email{low@ucar.edu}
\email{low@ucar.edu}
\affiliation{High Altitude Observatory, NSF National Center for Atmospheric Research, Boulder, Colorado, USA}
\author{S. W. McIntosh}
\email{smcintosh@lynker.com}
\affiliation{Lynker Space, Boulder, Colorado, USA}
\begin{abstract}

This self-contained theoretical study treats incompressible, free-boundary flows in a gravitating, ideal hydromagnetic star abutting vacuum, centered on the steady field-aligned flows of Chandrasekhar, Pendergast and Tsinganos, together with a novel family of steady cross-field flows, all as solutions of the axisymmetric Tsinganos equation.  In the absence of compressive waves and shocks, an incompressible fluid evolves by its frozen-in magnetic field propagating as transverse Alfv\`{e}n waves along the field lines, with pressure reacting instantly in place.  The origin of the steady flows rests on the Parker theory that everywhere-continuous flows are the exception rather than the rule because of a basic propensity for tangential field/flow discontinuities.  Astrophysical viscosity and electrical resistivity are not zero but are significant only over scales much smaller than macroscopic scales.  Such near-ideal fluids have the same propensity for tangential discontinuities but the near-discontinuities readily dissipate by small-scale, viscous-resistive magnetic reconnections.  The study treats the strictly ideal fluid separately in its own right, to construct a conceptual understanding of the turbulent creation of a steady flow in a self-organizing near-ideal fluid via irrepressible energy loss and field-topology changes as episodic reconnections run out of free energy.  The study suggests that metastable storage of steady vortices and twisted fields is a natural product of the Sun’s internal dynamo, to explain a recent, multi-instrument observation of solar-coronal eruptions persisting coherently in preferred longitudinal locations over solar-rotational timescales.\\
\center ({\it An HAO Study, Copyright \copyright 2026 by NSF NCAR.})
\\
\end{abstract}


\maketitle %\maketitle must follow title, authors, abstract and \pacs


\section{Introduction}

The formidable free-boundary problems posed by the steady ideal hydromagnetic flows of a given incompressible stellar mass $M_0$ abutting vacuum, fall into two classes depending on whether the field is wholly contained within the fluid or is extended across the stellar boundary into a vacuum potential field.  The former is energetically isolated, conserving its energy ${\mathcal E}$ as the sum of kinetic, magnetic, and gravitational potential energies.  Energy ${\mathcal E}$ is not conserved in the latter class by the loss of energy to Maxwell electromagnetic waves propagating away in the unbounded vacuum\citep{low1982}.  Here and elsewhere the unqualified terms of field, flux, helicity, and related entities shall mean the magnetic kind.  

The ideal fluid as a perfect electrical conductor deforms a wholly-contained field by bodily transporting the flux frozen in each fluid element.  Thus, field topology is invariant in a time-dependent flow, as is well known \citep{AlfvenFalthammar1963, LL1960, parker1979, kulsrud2005}.  A field is permanently identified by its invariant topology $\mathcal{T}$. Preserving $\mathcal{T}$ in the full freedom of motion in 3D space drives tangential discontinuities (TDs) in velocity ${\bf v}$ and field ${\bf B}$ to form spontaneously, as pointed out by Parker, Tsinganos and others \citep{parker1972, parker1994, jlp2010, syrovatskii1981, hk1985, ZweibelBoozer1985, low2007, low2015, low2019, low2023, tsinganos1981, tsinganos1984, zaslavskii1988}.   Anywhere in the fluid, vorticity and electric-current densities may intensify without limit in a layer of infinitesimal thickness, true to the physical meaning of zero viscosity and resistivity.  

There are two approaches to pose a steady free-boundary problem, the physical one seeking the steady flow of a specific fluid identified by its total mass $M_0$ and other permanent properties that include the invariant topology $\mathcal{T}$ of its embedded field.  That is, a fluid system is explicitly identified for one to construct its admissible steady states which may or may not contain TDs.  The steady-flow partial differential equations (PDEs) are not to be solved independently for they must couple to the magnetic-connectivity integrals defining a given $\mathcal{T}$. The other approach is mathematically simpler, seeking {\it analytic solutions} to the steady-flow PDEs subject to boundary conditions, a traditional approach in continuum mechanics that excludes TDs by the assumption of analyticity.  Each steady-flow solution so obtained identifies {\it in retrospect} the particular fluid the solution describes, including the frozen-in field of a specific topology $\mathcal{T}$ in the fluid.  The study will take the latter approach to obtain closed-form solutions to relate to the first approach, for a constructive physical understanding of the general free-boundary problem. 

The study is motivated by a unique dataset\citep{mcintosh2017} from the STEREO and SDO spacecraft observing the Sun’s tenuous outer atmosphere, the corona, during the period 2011-2013, that revealed persistent global-scale bands of magnetic activities meandering slowly in solar longitude.  This observation is suggestive of interior fluid vortices and twisted fields that are long-living and sporadically breaking across the solar surface to disrupt the local corona.   Assuming incompressibility, our theoretical study avoids the compressive sound and magneto-sonic waves in order to discover the simplest forms of hydromagnetic processes that correspond to their identifiable more complicated forms in the Sun.   

The physics of incompressible ideal steady stellar-flows is complete in its own right, providing a clear conceptual basis to be used later in the study for understanding what is meant by a near-ideal astrophysical fluid in which viscosity and electrical resistivity are weak, significant only over microscopic scales orders of magnitude smaller than astronomical macroscopic scales\citep{parker1979, low2015, low2019, low2023}.  The physical relationship between the ideal and the near-ideal fluids is a fundamentally interesting part of the study, suggesting that the emergent fluid vortices and twisted fields in the above coronal observation may be a natural product of the Sun’s interior dynamo\citep{parker1979}.   

Section II presents a derivation of the general, free-boundary, steady stellar flow.  Section III treats steady axisymmetric flows, illustrating general properties with the field-aligned flows of Chandrasekhar\citep{chandra1956, chandra1961}, Prendergast\cite{prendergast1956} and Tsinganos\cite{tsinganos1981}, and with a novel, cross-field flow as a solution of the Tsinganos equation\cite{tsinganos1981} subject to the stringent condition of Ferraro\cite{AlfvenFalthammar1963, ferraro1937}.  The cross-field flows are especially interesting, with ergodic flow and field lines intersecting endlessly on toroidal flux surfaces, the flow bodily and steadily transporting a conserved magnetic flux under the frozen-in condition.   Section IV presents several physical perspectives to view in broader physical context the fundamental properties of ideal fluids and their relationship with near-ideal fluids.  Section V concludes the study with a few remarks on future works.  The reference list of this article is not intended to be exhaustive, the list serving as a guide in our narrative.
   
\section{The steady free-boundary flows: general theory}

Consider the time-dependent equations describing a given mass $M_0$ of incompressible ideal fluid of uniform density $\rho_0$ flowing with velocity ${\bf v}$ and embedding a field ${\bf B}$:
\begin{eqnarray}
\label{momentum}
\rho_0 {\partial {\bf v} \over \partial t} + \rho_0 ( {\bf v} \cdot \nabla ) {\bf v}  &=& \frac{1}{4 \pi} \left( \nabla \times {\bf B} \right) \times {\bf B} - \nabla p  - \rho_0 \nabla U, \nonumber \\
\\
\label{solenoid_v}
\nabla \cdot {\bf v} &=& 0 , \\
\label{induction}
{\partial {\bf B} \over \partial t} &=& \nabla \times \left( {\bf v} \times {\bf B} \right) , \\
\label{solenoid_B}
\nabla \cdot {\bf B} &=& 0 , \\
\label{Newton}
\nabla^2 U &=& 4 \pi G \rho_0, 
\end{eqnarray}
\noindent 
in cgs units, with $p$, $U$ and $G$ respectively denoting fluid pressure, gravitational potential, and Newton's constant.  The fluid occupies a simply-connected domain $V$ of a fixed volume $V_0 = M_0/\rho_0$ with a deformable boundary $\partial V$ bordering vacuum.  

Solenoidal condition (\ref{solenoid_B}) on the field is redundant in time-dependent problems because if ${\bf B}$ is solenoidal at any one time, then ${\bf B}$ is solenoidal for all time by  induction equation (\ref{induction}).  Momentum equation (\ref{momentum}), incompressibility condition (\ref{solenoid_v}), induction equation (\ref{induction}), and Newton's gravitational equation (\ref{Newton}) pose 8 scalar PDEs for 8 scalar variables, namely, the respective three components of ${\bf v}$ and ${\bf B}$ plus the pair $(p, U)$.    

Taking the divergence operation across momentum equation (\ref{momentum}), the pressure everywhere is given at each instant of time by the Poisson equation
\begin{eqnarray}
\label{p_Poisson}
\nabla^2 p = &\nabla& \cdot \left\{\frac{1}{4 \pi} \left( \nabla \times {\bf B} \right) \times {\bf B} - \rho_0 {\partial {\bf v} \over \partial t} - \rho_0 ( {\bf v} \cdot \nabla ) {\bf v} \right\} \nonumber \\
&-& 4 \pi G \rho_0^2 ,  
\end{eqnarray}
\noindent
in terms of the fluid and field.  That is, pressure provides an instant reaction force that is not transmitted as waves. 

The following two equivalent forms of momentum equation (\ref{momentum}) are useful, each expressing inertial and Lorentz forces symmetrically.  Using the vector identity $\left( \nabla \times {\bf v} \right) \times {\bf v} = ( {\bf v} \cdot \nabla ) {\bf v} - \frac{1}{2} \nabla v^2$, the momentum equation takes the form:
\begin{eqnarray}
\label{momentum2}
{\partial {\bf v} \over \partial t} + \left( \nabla \times {\bf v} \right) \times {\bf v} &=& \frac{1}{4 \pi \rho_0} \left( \nabla \times {\bf B} \right) \times {\bf B} - \nabla P_B , \\ 
\label{Bernoulli_P}
P_B &=& {p \over \rho_0} + \frac{1}{2} v^2 + U ,
\end{eqnarray}
introducing Bernoulli pressure $P_B$ that involves the kinetic energy density.  Applying the above vector identity on ${\bf B}$ gives the other equivalent form, symmetrically expressing the  centrifugal and magnetic-tension forces,
\begin{eqnarray}
\label{momentum3}
{\partial {\bf v} \over \partial t} + \left( {\bf v} \cdot \nabla \right) {\bf v} &=& \frac{1}{4 \pi \rho_0} \left( {\bf B} \cdot \nabla \right) {\bf B} - \nabla P_T , \\ 
\label{Total_P}
P_T&=& {1 \over \rho_0} \left[ p + \frac{B^2}{8 \pi} \right] + U ,
\end{eqnarray}
introducing the total pressure $P_T$ that involves the magnetic energy density.  

Rewrite momentum equation (\ref{momentum2}) in terms of vorticity ${\bf w}$, 
\begin{eqnarray}
\label{vorticity_w} 
{\bf w} = \nabla \times {\bf v} , ~~~~~~~~~~~~~~~~~~\\
\label{momentum4}
{\partial {\bf w} \over \partial t} + \nabla \times\left( {\bf w} \times {\bf v} - \frac{1}{4 \pi \rho_0} \left( \nabla \times {\bf B} \right) \times {\bf B} \right) = 0 .
\end{eqnarray}
Equations (\ref{induction}), (\ref{vorticity_w}) and (\ref{momentum4}) are a hyperbolic system of PDEs describing incompressible Alfv\`{e}n waves, the waves capable of deforming the free boundary $\partial V$.  The Alfv\`{e}n-waves propagate at phase-velocity ${\bf v}_A = {\bf B}(4 \pi \rho_0)^{-\frac{1}{2}}$ along field lines that are the space-time characteristic curves\citep{ch1960, schwartz1966, cf1976} of the PDEs.  That is, the Alfv\`{e}n-waves nonlinearly define their respective wave-paths as they propagate.  Converging or merging characteristics readily result in the dependent variables, integrated along the characteristics, becoming multivalued in space beyond a critical time, a fundamental behavior of nonlinear hyperbolic PDEs.  The evolution continues physically with the fluid and field remaining single valued in space by the formation of tangential discontinuities (TDs) in $p$, ${\bf v}$ and ${\bf B}$ that move bodily as fluid surfaces.  

In a compressible fluid, magneto-sonic shock waves as discontinuities arise nonlinearly by the convergence or merging of the space-time hyperbolic characteristic curves along which magneto-sonic waves propagate\citep{LL1960}.  In this case, the fluid passing through the shocks are necessarily dissipative and irreversible, generating specific amounts of entropy created by viscosity and electrical resistivity implied by the conservation laws governing the shocks.  Quite distinct, the formation of a magnetic TD is a reversible process in the ideal incompressible fluid, with new TDs forming and existing TDs reversibly dissolving back into the global flow.  

Induction equation (\ref{induction}) is an Eulerian description of the flow and field at each space-time point without involving where each fluid element is located at any given time.  The equivalent Lagrangian description carries more information.  Under equation (\ref{induction}), the flux ${\mathcal F}(\Sigma)$ across each fluid surface $\Sigma$ is constant in time,
\begin{equation}
\label{frozen-in}
{d\over dt} {\mathcal F}(\Sigma)  \equiv {d\over dt} ~ \int_{\Sigma}~{\bf B} \cdot {\bf \hat n} ~d\Sigma = 0 ,  
\end{equation}
\noindent   
where $d\Sigma~{\bf \hat n}$ denotes a directed area element.  A flux surface is permanently identified by the same fluid elements on it.  The textbook definition of topology suffices for our discussion, topological properties being those of the frozen-in flux surfaces and field lines that are invariant under all continuous fluid deformations.  In the absence of viscosity and electrical resistivity, two ideal fluid parcels can slip tangentially and discontinuously along a common flux-surface boundary where a vortex sheet and an electric-current sheet form, both sheets infinitesimally thin.

\subsection{The steady free-boundary flows}

A perfect fluid observes zero electric field in the rest frame of each fluid element.  By the non-relativistic transformation of electro-magnetic fields, the electric field in the laboratory frame is
\begin{equation}
\label{electric_field1}
{\bf E} = - {1 \over c} \left( {\bf v} \times {\bf B} \right) , 
\end{equation}
\noindent
defined by ${\bf v}$ and ${\bf B}$ as solutions to the hydromagnetic equations.  The speed of light $c$ arises from the use of cgs units, no relativistic effect implied.  In non-relativistic description, equation (\ref {electric_field1}) shows that  $c{\bf E}$ is of the same leading order of magnitudes as the vector product of velocity and field.   Induction equation (\ref{induction}) expresses the Maxwell equation
\begin{equation}
\label{Faraday}
{1 \over c}{\partial {\bf B} \over \partial t} + \nabla \times {\bf E} = 0 . 
\end{equation}
\noindent
In the laboratory frame, this equation describes the dynamo generation of fresh magnetic flux such that, under perfect electrical conductivity, the field behaves as frozen into the fluid.  

With the preceding clarification on the electric field, a steady flow is governed by the electric field being potential:
\begin{eqnarray}
\label{force-balance1}
\left( \nabla \times {\bf v} \right) \times {\bf v} &=& \frac{1}{4 \pi \rho_0} \left( \nabla \times {\bf B} \right) \times {\bf B} - \nabla P_B , \\ 
\label{potential_E}
\nabla \times {\bf E} &=& 0 ;
\end{eqnarray}
\noindent
see momentum equation (\ref{momentum2}) and Faraday equation (\ref{Faraday}).   Bernoulli pressure $P_B$ and electric field ${\bf E}$ are given by equations (\ref{Bernoulli_P}) and (\ref{electric_field1}), respectively, both ${\bf v}$ and ${\bf B}$ being solenoidal as described by equations (\ref{solenoid_v}) and (\ref{solenoid_B}).

We focus attention on two general classes of steady flows, the first being the field-aligned flows 
\begin{equation}
\label{field-aligned}
{\bf v} = \gamma_0 {{\bf B} \over 4 \pi \rho_0},  
\end{equation}
\noindent
$\gamma_0$ a given constant, for which ${\bf E} = 0$.  The other class comprises cross-field flows subject to the Ferraro\citep{ferraro1937} iso-rotation condition expressed by the presence of a potential electric field $c {\bf E} = -\nabla W$,
\begin{equation}
\label{cross-field}
{\bf v} \times {\bf B} = \nabla W .  
\end{equation}
\noindent
The field-aligned flows may be viewed as cases of the cross-field flows in the limit $W \rightarrow 0$.    

The steady-flow equations  are subject to the stress-free conditions governing the TDs in the global flow and at the free boundary $\partial V$ abutting vacuum.  There are two types of free boundary conditions at $\partial V$, the first describing ${\bf B}$ wholly contained in $V$ and tangential along $\partial V$.  Let us use the steady force-balance expressed by 
\begin{equation}
\label{force_balance2}
\left( {\bf v} \cdot \nabla \right) {\bf v} = \frac{1}{4 \pi \rho_0} \left( {\bf B} \cdot \nabla \right) {\bf B} - \nabla P_T ,
\end{equation}
\noindent
setting time derivatives to zero in momentum equation (\ref{momentum3}).  Integrating equation (\ref{force_balance2}) across $\partial V$ into vacuum, we require the total pressure $P_T$ to be continuous because the magnetic-tension and centrifugal forces have no contribution.  The gravitational potential energy $U$ being continuous also has no contribution.  Therefore, the sum $p + {B^2 \over 8 \pi}$ of two non-negative quantities, must be continuous across $\partial V$, giving the boundary conditions 
\begin{equation}
\label{stress-free_condition1}
p = |{\bf B}| = 0 ~~~\mathrm{at} ~~ {\partial V} ,
\end{equation}
\noindent
and permitting ${\bf v}$ to be finite at $\partial V$ and falling discontinuously to zero in vacuum.  

If, instead, the interior field ${\bf B}$ extends across $\partial V$ into a vacuum potential field ${\bf B}_{pot}$, the two fields must be continuous at the free boundary.  The boundary normal components of the two fields are continuous under the solenoidal condition.  Whereas, their tangential components must also be continuous avoid a TD with a discrete Lorentz force along $\partial V$ not possible to be balanced by the fluid.  The normal component of the fluid velocity must vanish at $\partial V$ under the solenoidal condition.  Whereas, the fluid velocity may have a non-zero component along $\partial V$, bodily and steadily moving the interior field ${\bf B}$ and its connected exterior potential field ${\bf B}_{pot}$.  With these considerations, the following boundary conditions are imposed: 
\begin{equation}
\label{stress-free_condition2}
p =  0,  ~~~{\bf B} = {\bf B}_{pot} , ~~~\mathrm{at} ~~ {\partial V} ,
\end{equation}
\noindent
permitting the tangential ${\bf v}$ dropping discontinuously to zero in the vacuum exterior.

In both wholly contained fields and fields connected to their respective external potential fields, the stellar flow may contain TDs in ${\bf v}$ and ${\bf B}$ in its interior.  The force balance at each interior TD then takes the form of continuity of the total pressure:
\begin{equation}
\label{stress-free_condition3}
\Delta \left[ p + {B^2 \over 8 \pi} \right]_{TD} = 0 ,
\end{equation}
\noindent
$\Delta$ denoting the difference in total pressure.  No condition applies to the tangential velocity ${\bf v}$ on the two sides of a TD. 

The free-boundary conditions (\ref{stress-free_condition1})-(\ref{stress-free_condition3}) are not posed for an {\it a priori} given boundary.  The shapes of the boundary $\partial V$ and interior TDs, if any exist, are among the mathematical unknowns in the free-boundary problem, to be determined self-consistently with a solution of the governing steady-flow PDEs.  Note that although the gravitational potential $U$ satisfies the simple linear Poisson equation (\ref{Newton}), $U$ must be determined self-consistently with the shape of $\partial V$ as a part of the mathematical solution.  

Kinetic and magnetic energies, as well as the incompressible pressure, are all expansive.  Although the magnetic tension force if inward directed has a confining effect, it can never confine its corresponding magnetic pressure force\cite{chandra1961}.  Gravity is the only means of holding the star against the expansion forces or against the incompressible fluid breaking up into separate volumes.  Through $\partial V$ as an unknown, the coupling of Newton’s equation with the other hydromagnetic equations determine the self-confining gravitational potential $U$.

\subsection{Foliation of 3D flux surfaces and ergodic field lines}

Given an analytic field ${\bf B}(r, \theta, \varphi)$ in spherical coordinates $(r, \theta, \varphi)$, its field lines are described by the system of ordinary differential equations (ODEs):
\begin{equation}
\label{FLeqns}
{dr \over B_r} = {r d\theta \over B_{\theta}} ={r \sin\theta d\varphi \over B_{\varphi}} ={d\ell \over |{\bf B}|} ,
\end{equation}
\noindent
where $\ell$ is the field-line path length defined by $d\ell^2 = dr^2 + r^2 d\theta^2 + r^2 \sin^2 \theta d\varphi^2$.  Coordinate $\varphi$ may be chosen as an independent variable along any field line, with two first order ODEs for dependent variables $r(\varphi)$ and $\theta(\varphi)$ along the field line identified by an “initial condition” $\varphi = \varphi_0$, a constant.  There are two integration constants determined by $\left[ r_0 = r(\varphi_0), \theta_0 = \theta(\varphi_0) \right]$ locating a particular point $(r_0, \theta_0, \varphi_0)$ on the line.  

For an arbitrarily prescribed 3D varying field, meaning ${\bf B}(r, \theta, \varphi)$ is an arbitrarily prescribed analytic function of space, its field lines generally include endless ergodic lines that do not cross under the solenoidal condition.  A single ergodic line can {\it fill} an entire flux surface or magnetic sub-volumes bounded by flux surfaces.  Filling is taken in the limiting sense of the endless ergodic line approaching as close as desired to any point in its filled surface or volume.  Field-lines are conceptually unambiguous physical objects uniquely defined by ODEs (\ref{FLeqns}).    

A field line is locally the intersection of a freely selected pair of flux surfaces, but this geometric relationship cannot be global for ergodic field lines.  An ergodic field line filling up a volume of space would mean that the two flux surfaces intersecting along the line also fill up the volume of space.  This seemingly absurd geometric property does not physically negate the existence of ergodic field lines and their flux surfaces.  It only means that there is no global mathematical representation of the ergodic field lines and their flux surfaces, with ODEs (\ref{FLeqns}) as the mathematical basis to connect the field lines and their flux surfaces found in distinct adjacent spatial regions. 

Given ${\bf B}(r, \theta, \varphi)$, its field lines in a sufficiently small volume of space may be identified as the mutual intersections between two independent families of flux surfaces described as the level surfaces of two scalar functions $\xi_1(r, \theta, \varphi)$ and $\xi_2(r, \theta, \varphi)$, hereafter the Euler potentials.  It follows that ${\bf B} = B_0(\xi_1, \xi_2) \nabla \xi_1 \times \nabla \xi_2$, with amplitude $B_0(\xi_1, \xi_2)$ being a constant along each field line to satisfy the solenoidal condition. 

Identifying the field lines as the mutual intersections between the $\xi_1$ and $\xi_2$ flux surfaces is not unique, of course.  Given $(\xi_1, \xi_2)$, $\left[\Xi_1(\xi_1, \xi_2), \Xi_2(\xi_1, \xi_2)\right]$ is another pair of Euler potentials describing the same physical field line to be lying in constant-$\Xi_1$ and constant-$\Xi_2$ flux surfaces of ${\bf B}(r, \theta, \varphi)$, where $(\Xi_1, \Xi_2)$ are freely prescribed continuous functions of two variables with their corresponding field amplitude $B_0(\Xi_1, \Xi_2)$ suitably re-defined.  The mathematical freedom to exhaustively order the field lines into two independent families of Euler potentials allows one, with no loss of generality, to choose a pair $(\xi_1, \xi_2)$ such that $B_0 = 1$ and the given field is then represented simply as
\begin{equation}
\label{Euler_rep}
{\bf B} = \nabla \xi_1 \times \nabla \xi_2 .
\end{equation}
\noindent
This field representation is local, the geometric absurdity of a single line filling a flux surface or a magnetic volume is reconciled by the realization that a global flux system generally contains ergodic sub-volumes, each sub-volume containing a single ergodic line filling it.   The sub-volume is contiguous with layers of globally defined flux surfaces, each surface possibly also containing a single ergodic line.  A field line has zero cross-section, only having a finite length if it is spatially isolated to close upon itself.  The Euler representation breaks down in regions of space that include ergodic field lines for which one or both $\xi_1$ and $\xi_2$ flux-surfaces cannot be globally defined, although, to emphasize, the ergodic field lines and their flux surfaces are each conceptually well defined.  There is much to learn about the topology of field lines by constructing explicit examples.  

\subsection{Chandrasekhar and Tsinganos-Prendergast field-aligned flows}

Mathematically trivial but physically far reaching, force-balance equation (\ref{force-balance1}) is satisfied for $\gamma_0 = \pm 1$ for {\it all} velocities set parallel or anti-parallel to the Alfv\`{e}n velocity in an {\it arbitrarily} prescribed ${\bf B}$, the Lorentz and fluid inertial forces balancing each other with the Bernoulli pressure $P_B$ and the total pressure $P_T$ being identical and uniform in space, given by a constant $P_0$:
\begin{eqnarray}
\label{Chandrasekhar1}
{\bf v} &=& \pm \frac{ {\bf B}}{\sqrt{4 \pi \rho_0}} , \\
\label{Chandrasekhar2}
P_B &\equiv& P_T \nonumber = {1 \over \rho_0}\left(p + {|{\bf B}|^2 \over 8 \pi}\right)  + U  = P_0 .
\end{eqnarray}
\noindent
These are the Chandrasekhar equipartition flows\citep{chandra1956, chandra1961, tsinganos1981} with the kinetic energy density $\frac{1}{2} \rho_0 v^2$ equal to the field energy density$\frac{1}{8 \pi} B^2$ everywhere.  Magnetic fields propagate as nonlinear waves along the field at the Alfv\`{e}nic wave velocity $\frac{ {\bf B}}{\sqrt{4 \pi \rho_0}}$ in the incompressible fluid.  In the equipartition flows, the field is stationary, with no free wave-energy, in the rest frame of every parcel of fluid, and is demonstrably linearly stable\citep{chandra1956, chandra1961}.  

The equipartition flows define a manifold of stable Chandrasekhar incompressible steady stars.  Each star of mass $M_0$ and uniform density $\rho_0$ occupies a volume $V$ of a fixed volumetric measure $V_0 = M_0/\rho_0$ bounded by a stress-free boundary $\partial V$ whose shape defines the self-confining gravitational force.  Chandrasekhar stars with wholly contained 3D fields are readily constructed by a mere prescription of the field to fit a geometric shape of $\partial V$ subject to the field and its aligned flow vanishing at the stellar boundary. 

\subsection{Steady cross-field flows}

Steady cross-field flows under the frozen-in condition in 3D space are subject to Ferraro’s severe condition that each individual fluid element must carry its frozen flux in its steady motion with no change to the steady distribution of the field.  Such flows are expected to be composed of regions of continuous flow bounded by interior stress-free magnetic TDs as fluid boundaries. Here we take an overview of the general governing steady equations, representing the solenoidal ${\bf v}$ and ${\bf B}$ locally by their pairs of Euler potentials, to keep in mind in the study of a novel family of axisymmetric cross-field flows in Section III.    

Denote a point in 3D Cartiesian space by the coordinate vector ${\bf x}$ and express solenoidal ${\bf v}$ and ${\bf B}$ by the pairs of Euler potentials $[\Phi({\bf x}), W_1({\bf x})]$ and $[A({\bf x}), W_2({\bf x})]$, respectively,  
\begin{eqnarray}
\label{v_PhiW1}
{\bf v} &=& \nabla \Phi \times \nabla W_1 = \nabla \times \left[ \Phi \nabla W_1 \right] , \\ 
\label{B_AW2} {\bf B} &=& \nabla A \times \nabla W_2 = \nabla \times\left[ A \nabla W_2 \right] . 
\end{eqnarray}
\noindent 
The velocity-lines lie on two distinct families of flow surfaces of constant $\Phi$ and constant $W_1$.  To reword, the two families of surfaces mutually intersect along the velocity-lines.  Similarly, the mutual intersections of the two families of constant-$A$ and constant-$W_2$ are the field-lines.   Ferraro’s condition (\ref{cross-field}) is met by the velocity-lines and field-lines lying on surfaces of constant electrostatic potential $W$.  Therefore, we may identify the flow-surfaces of constant $W_1$ and the flux-surfaces of constant $W_2$ with the electric-potential surface of constant $W$, i.e., $W_1 \equiv W_2 \equiv W$,
\begin{eqnarray}
\label{v_PhiW}
{\bf v} &=& \nabla \Phi \times \nabla W \equiv \nabla \times \left[ \Phi \nabla W \right] , \\ 
\label{B_AW} {\bf B} &=& \nabla A \times \nabla W \equiv \nabla \times\left[ A \nabla W \right] , 
\end{eqnarray}
\noindent 
by which Ferraro’s condition (\ref{cross-field}) takes the form
\begin{equation}
\label{electric_field3}
\left( \nabla \Phi \times \nabla A \right) \cdot \nabla W = 1 ,
\end{equation}
\noindent
and the following can be evaluated
\begin{eqnarray}
\label{vorticityXv}
\left( \nabla \times {\bf v} \right) \times {\bf v} = \left\{ \nabla \cdot \left[ \left( \nabla W \cdot \nabla \Phi \right) \nabla W \right] - \nabla \cdot \left[ |\nabla W|^2 \nabla \Phi \right] \right\} \nabla \Phi \nonumber  \\ 
+ \left\{ \nabla \cdot \left[ \left( \nabla W \cdot \nabla \Phi \right) \nabla \Phi \right] - \nabla \cdot \left[ |\nabla \Phi|^2 \nabla W \right] \right\} \nabla W , ~~~~~~~~~~~~~ \\ 
\label{LorenzForce} 
\left( \nabla \times {\bf B} \right) \times {\bf B} = \left\{ \nabla \cdot \left[ \left( \nabla W \cdot \nabla A \right) \nabla W \right] - \nabla \cdot \left[ |\nabla W|^2 \nabla A \right] \right\} \nabla A \nonumber \\ 
+ \left\{ \nabla \cdot \left[ \left( \nabla W \cdot \nabla A \right) \nabla A \right] - \nabla \cdot \left[ |\nabla A|^2 \nabla W \right] \right\} \nabla W , ~~~~~~~~~~~~~
\end{eqnarray}
\noindent 
where we have used the vector identities $\nabla \cdot \left( {\bf a} \times {\bf b}\right) \equiv {\bf b} \cdot \nabla \times {\bf a} - {\bf a} \cdot \nabla \times {\bf b}$ and $\nabla \times \left[\nabla f({\bf x}) \right] \equiv 0$.  

We may treat $[\Phi({\bf x}), A({\bf x}), W({\bf x})]$ to be the curvilinear-coordinates of a point located at Cartesian ${\bf x} = (x_1, x_2, x_3)$ in 3D space.  Then, Ferraro’s condition (\ref{electric_field3}) has the simple geometric interpretation
\begin{eqnarray}
d\Phi ~dA ~dW &\equiv& \left( \nabla \Phi \times \nabla A \right) \cdot \nabla W dx_1 dx_2 dx_3 \nonumber \\
&=& dx_1 dx_2 dx_3 ,
\end{eqnarray}
\noindent
that the transformation $(x_1, x_2, x_3) \rightarrow (\phi, A, W)$ is volume preserving.  Taking the Bernoulli pressure $P_B({\bf x}) \equiv P_B\left[ \Phi({\bf x}), A({\bf x}), W({\bf x}) \right]$, expressed in terms of the curvilinear coordinates $(\Phi, A, W)$ at each point in space, we have the gradient 
\begin{equation}
\label{grad_PB}
\nabla P_B = {\partial P_B \over \partial \Phi} \nabla \Phi + {\partial P_B \over \partial A} \nabla A + {\partial P_B \over \partial W} \nabla W . 
\end{equation}
\noindent
Resolving force balance equation (\ref{force-balance1}) into components in the three directions of the independent gradients $\left[ \nabla \Phi, \nabla A, \nabla W \right]$ and summing the components in each of the directions to zero, we obtain the following second-order nonlinear PDEs describing the steady-flow force balance:
\begin{eqnarray}
\label{FB_Phi}
\nabla \cdot \left[ \left( \nabla W \cdot \nabla \Phi \right) \nabla W \right] - \nabla \cdot \left[ |\nabla W|^2 \nabla \Phi \right]&& \nonumber \\
 + {\partial \over \partial \Phi} P_B\left(\Phi, A, W \right) = 0 ,&&\\ 
\label{FB_A}
\nabla \cdot \left[ \left( \nabla W \cdot \nabla A \right) \nabla W \right] - \nabla \cdot \left[ |\nabla W|^2 \nabla A \right]&& \nonumber \\ + {\partial \over \partial A} P_B\left(\Phi, A, W \right) = 0 , &&\\
\label{FB_W}
\nabla \cdot \left[ \left( \nabla W \cdot \nabla \Phi \right) \nabla \Phi \right] -\nabla \cdot \left[ \left( \nabla W \cdot \nabla A \right) \nabla A \right] && \nonumber \\ + \nabla \cdot \left[ \left(|\nabla A|^2 - |\nabla \Phi|^2 \right) \nabla W \right] + {\partial \over \partial W} P_B\left(\Phi, A, W \right) &= & 0 . \nonumber \\
\end{eqnarray}
\noindent 
Coupled to equation (\ref{electric_field3}), we have a compete set of four PDEs to be solved for $\left(\Phi, A, W, P_B \right)$ describing a steady flow subject to the boundary conditions for internal TDs and the free-boundary of an incompressible star.  The Bernoulli pressure $P_B$ carries the fluid’s self-gravity which is the only attractive force that holds the steady flow and field in the incompressible star.

These highly nonlinear PDEs are new, not mathematically understood, and complicated by the fact that Euler potentials and the flux surfaces they represent are generally not definable globally in a 3D-varying fluid, a fundamental point to keep in mind in the study of axisymmetric flows. 

\section{Axisymmetric Tsinganos steady-flows} 

With ${\partial \over \partial \varphi} \equiv 0$ in spherical coordinates $(r, \theta, \varphi)$, the solenoidal vectors ${\bf v}$ and ${\bf B}$ are each the superposition of a poloidal field and a toroidal field,
\begin{eqnarray}
\label{phi_Psi}
{\bf v} &=& {\bf v}_p + {\bf v}_t , \nonumber \\
&=& v_0 \nabla \phi \times \nabla \varphi + v_0 \Psi \nabla \varphi \nonumber \\
&=& {v_0 \over r  \sin \theta} \left( {1 \over r}  {\partial \phi \over \partial \theta}, -{\partial \phi \over \partial r}, \Psi \right), \\
\label{A_Q}
{\bf B} &=& {\bf B}_p + {\bf B}_t , \nonumber \\
&=& B_0 \nabla A \times \nabla \varphi + B_0 Q \nabla \varphi \nonumber \\
&=&  {B_0 \over r  \sin \theta} \left( {1 \over r}  {\partial A \over \partial \theta}, -{\partial A \over \partial r}, Q \right),
\end{eqnarray}
\noindent 
where $\nabla \varphi = {\hat \varphi \over r \sin \theta}$, and $\left(v_0, B_0\right)$ are normalization constants.  The flow/field lines of a star are respectively projected onto the $r$-$\theta$ planes as {\it closed} contours of constant poloidal stream/flux functions $\left[\phi(r, \theta), A(r, \theta)\right]$.  In 3D space, the level surfaces of $\left[\phi(r, \theta), A(r, \theta)\right]$ are nested toroidal surfaces on which the flow/field lines stream with toroidal $\varphi$-displacements dictated by $v_{\varphi} = v_0 {\Psi \over r \sin \theta}$ and $B_{\varphi} = B_0 {Q \over r \sin \theta}$.  

\subsection{The Tsingnos PDE}
  
We re-derive the axisymmetric Tsinganos\cite{tsinganos1981} steady-flow PDE.  The vorticity ${\bf w} = \nabla \times {\bf v}$ and current density ${\bf J}={4 \pi \over c}\nabla \times{\bf B}$ are both solenoidal, each also the sum of a poloidal and toroidal part:
\begin{eqnarray}
\label{vorticity}
\nabla \times {\bf v} &=& v_0 \nabla \varphi \times \nabla \Psi - v_0 {\mathcal L} \phi \nabla \varphi \nonumber \\
 &=&{v_0 \over r  \sin \theta} \left( {1 \over r}  {\partial \Psi \over \partial \theta}, -{\partial \Psi \over \partial r}, - {\mathcal L} \phi \right) , \\
\label{currentden}
\nabla \times{\bf B} &=& B_0 \nabla \varphi \times \nabla Q - B_0 {\mathcal L} A \nabla \varphi \nonumber \\
&=& {B_0 \over r  \sin \theta} \left( {1 \over r}  {\partial Q \over \partial \theta}, -{\partial Q \over \partial r}, - {\mathcal L}A \right) , \\
\label{operatorL}
{\mathcal L} & \equiv & {\partial^2 \over \partial r^2} + {1 - \mu^2 \over r^2} {\partial^2 \over \partial \mu^2} ; ~~~\mu = \cos \theta .
\end{eqnarray}  
A surface of constant-$A$, or constant-$\phi$, is the boundary of a toroidal tube of twisted flux, or a channel of vortical fluid flow, respectively encircling the symmetry axis.   The steady flow equations (\ref{force-balance1}) and (\ref{cross-field}) then give:
\begin{eqnarray}
\label{fbalance1}
v_0^2 {\mathcal L} \phi \nabla \phi - v_A^2 {\mathcal L}A \nabla A + \frac{1}{2}\nabla \left( v_0^2 \Psi^2 - v_A^2 Q^2  \right)&& \nonumber \\
+ \left[ v_0^2 {1 \over r} {\partial \left( \Psi, \phi \right) \over \partial \left( r, \theta \right)} - v_A^2 {1 \over r} {\partial \left( Q, A \right) \over \partial \left( r, \theta \right)}\right] {\bf \hat \varphi} + R^2 \nabla P_B = 0 , &&~~~~~\\
\label{Ferraro1}
\Psi \nabla A - Q \nabla \phi + {1 \over r} {\partial \left( \phi, A \right) \over \partial \left( r, \theta \right)} {\bf \hat \varphi} + {c ~r^2 \sin^2 \theta \over v_0v_A \sqrt{4 \pi \rho_0}} \nabla W = 0 , &&~~
\end{eqnarray}
\noindent
introducing $R^2 = r^2 \sin^2 \theta$ and the constant Alfv\`{e}n speed $v_A = {B_0 \over \sqrt{4 \pi \rho_0}}$.  

The gradients in equation (\ref{Ferraro1}) have no component in the ${\hat \varphi}$ direction, and it follows that
\begin{equation}
\label{phi(A)}
{\partial \left( \phi, A \right) \over \partial \left( r, \theta \right)} ~\equiv~ {\partial \phi \over \partial r}{\partial A \over \partial \theta} -  {\partial A \over \partial r}{\partial \phi \over \partial \theta} = 0  ~~\Leftrightarrow~~ \phi(r, \theta) = \phi[A(r, \theta)] , 
\end{equation}
\noindent
requiring the constant-$A$ flux surfaces and constant-$\phi$ flow surfaces to coincide, describing the aligned poloidal flow and field,
\begin{eqnarray}
\label{v_p}
{\bf v}_p &=& v_0 {d\phi \over dA} \nabla \varphi \times \nabla A , \\
\label{B_p}
{\bf B}_p &=& B_0\nabla \varphi \times \nabla A .
\end{eqnarray}
\noindent
Equation (\ref{Ferraro1}) now reads
\begin{equation}
\label{Ferraro2}
\left(\Psi  - Q {d\phi(A) \over dA}\right) \nabla A + {c~r^2 \sin^2 \theta \over v_0v_A \sqrt{4 \pi \rho_0}} ~\nabla W = 0 ,
\end{equation}
\noindent
requiring that $W(r, \theta) = W[A(r, \theta)]$ so that
\begin{equation}
\label{QPsiW}
Q{d\phi \over dA} - \Psi  = {c~r^2 \sin^2 \theta \over v_0v_A\sqrt{4 \pi \rho_0}} ~ {dW(A) \over dA} .
\end{equation}
\noindent
Three sets of level surfaces coincide geometrically, namely, the constant-$A$ flux surfaces, constant-$\phi(A)$ flow surfaces and constant-$W(A)$ electrostatic potential surfaces, an expression of  the Ferraro iso-rotational condition.  

Force-balance equation (\ref{fbalance1}) demands that the magnetic-tension and centrifugal forces must sum to zero in the $\varphi$-direction since gradients have no components in that direction.  Subject to 
$\phi (r, \theta) = \phi [A(r, \theta)]$, the vanishing of the sum of the two forces in the $\varphi$ direction is expressed by a scalar function $K(r, \theta)$ as follows,  
\begin{eqnarray}
\label{varphi_force}
v_0^2 {1 \over r} {\partial \left( \Psi, \phi \right) \over \partial \left( r, \theta \right)} - v_A^2 {1 \over r} {\partial \left( Q, A \right) \over \partial \left( r, \theta \right)} = &0& \Rightarrow {\partial \left( K, A \right) \over \partial \left( r, \theta \right)} = 0 \nonumber \\
~\Rightarrow ~K (r, \theta) &=& K\left[A(r, \theta)\right] ,
\end{eqnarray}
\noindent
with $K(A)$ given in terms of $Q(r, \theta)$, $\Psi(r, \theta)$ and $\phi(A)$: 
\begin{equation}
\label{QPsiOmega}
v_A^2 Q - v_0^2 \Psi {d\phi \over dA} = K\left[A(r, \theta)\right] .
\end{equation}
\noindent
The $\varphi$-components of the magnetic-tension and centrifugal forces, in general, are not zero but must mutually balance, expressed by equation (\ref{QPsiOmega}) with $K(r, \theta)$, hereafter the $\varphi$-force integral, taking a constant value on each constant-$A$ flux surface.  

We may set $v_0 = v_A$ with no loss of generality since the function $\phi (r, \theta) = \phi [A(r, \theta)]$ defines the relative strengths of the two poloidal vectors ${\bf B}_p$ and ${\bf v}_p$.   Using this normalization, each Tsinganos steady flow comprises the aligned poloidal pair $[{\bf B}_p, {\bf v}_p]$ and the toroidal pair $[{\bf B}_t, {\bf v}_t]$, the latter given in terms of $\Psi$ and $Q$ as the solutions of the linear algebraic equations (\ref{QPsiW}) and (\ref{QPsiOmega}):
\begin{eqnarray}
\label{PsiQ1}
\Lambda(A) \Psi &=& K (A) {d\phi \over dA} - {c ~ r^2 \sin^2 \theta \over \sqrt{4 \pi \rho_0}}  {dW(A) \over dA} , \\ 
\label{PsiQ2}
\Lambda(A) Q &=&K (A) - {c  ~ r^2 \sin^2  \theta \over \sqrt{4 \pi \rho_0}} {d\phi \over dA} {dW(A) \over dA} ; \\
\label{PsiQ3}
\Lambda(A) &=& v_A^2 \left[ 1- \left({d\phi \over dA} \right)^2 \right] ,
\end{eqnarray}
\noindent
each steady flow generated by prescribing the free functions $\phi(A), W(A), K(A)$.  The factor $\Lambda(A)$ is a measure of the difference in energy density between the poloidal field and the aligned poloidal velocity:  
\begin{eqnarray}
\label{supersub-Alfven}
\Delta {\mathcal E}  &=& {B_p^2 \over 8 \pi} - \frac{1}{2} \rho_0 v_p^2 \nonumber \\
&=& \frac{1}{2}\rho_0\Lambda(A) |\nabla \varphi \times \nabla A|^2 .
\end{eqnarray}
\noindent
Along the flux surfaces identified by $\Lambda(A) = 0$, the poloidal part of the steady flow is Alfv\`{e}nic, meaning $|{\bf v}_p| = {{\bf B}_p| \over \sqrt{4 \pi \rho_0}}$.  The poloidal flow changes from sub-Alfv\`{e}nic $\left[\Lambda(A) > 0\right]$ to super-Alfv\`{e}nic $\left[\Lambda(A) < 0 \right]$ across each critical $\Lambda(A) = 0$ surface.  Nonlinear Alfv\`{e}n waves traveling along field lines behave differently on the two sides of this critical surface.  Magnetic disturbances can pervade throughout a strictly sub-Alfv\`{e}nic flow but would readily pile up in a forward super-Alfv\`{e}nic flow, an interesting unexplored property.  The distinct behaviors of the front of a packet of nonlinear Alfv\`{e}n waves, on the two sides of a $\Lambda(A) = 0$ surface in a steady flow suggest that this critical flux surface would become a magnetic TD.     

With no loss of generality, redefine 
\begin{eqnarray}
\label{PsiQ4}
K (A) &=& \Lambda(A) S_1(A) , \\ 
\label{PsiQ5}
{dW(A) \over dA} &=& \Lambda(A) S_2(A) ,
\end{eqnarray}
\noindent
in terms of free functions $S_1(A)$ and $S_2(A)$, rewriting equations (\ref{PsiQ1}) and (\ref{PsiQ2}) as
\begin{eqnarray}
\label{PsiQ6}
\Psi &=&  {d\phi \over dA} S_1 (A)- {c~r^2 \sin^2 \theta \over \sqrt{4 \pi \rho_0}}  S_2 (A) , \\ 
\label{PsiQ7}
Q &=& S_1 (A) - {c~r^2 \sin^2 \theta \over \sqrt{4 \pi \rho_0}} {d\phi \over dA} S_2 (A) . 
\end{eqnarray}
\noindent 
A straightforward algebra then gives the following tidy term
\begin{eqnarray}
\label{Q_Psi}
{\mathcal S}(A, R) &=&v_A^2 \left( Q^2 - \Psi^2 \right) \nonumber \\
&=&  \Lambda(A) \left[S^2_1(A) - {c^2 R^4 \over 4 \pi \rho_0} S^2_2(A)\right] , 
\end{eqnarray}
\noindent
with $R = r \sin \theta$.  Force-balance equation (\ref{fbalance1}) can then be rewritten as the vanishing of the sum of three gradients:
\begin{eqnarray}
\label{tsing_0}
\left[\Lambda(A) {\mathcal L} A - v_A^2{d\phi \over dA} {d^2\phi \over dA^2} |\nabla A|^2 \right] \nabla A +\frac{1}{2} \nabla {\mathcal S}(A, R)&&  \nonumber \\
+ R^2 \nabla P_B = 0 , &&
\end{eqnarray}
\noindent
with $\phi(r, \theta) = \phi[A(r, \theta)]$.

Resolving these three gradients into components in the two independent directions of $\nabla A$ and $\nabla R$, equation (\ref{tsing_0}) takes the form:
\begin{eqnarray}
\label{tsing_1}
\left[\Lambda(A) {\mathcal L} A - v_A^2 {d\phi \over dA} {d^2\phi \over dA^2} |\nabla A|^2 + \frac{1}{2} {\partial \over \partial A} {\mathcal S} (A, R) \right. \nonumber \\
\left. + R^2 {\partial \over \partial A} P_B(A, R) \right] \nabla A \nonumber \\
+ \left[\frac{1}{2} {\partial \over \partial R} {\mathcal S} (A, R) + R^2 {\partial \over \partial R}  P_B(A, R) \right] \nabla R = 0 ,
\end{eqnarray}
\noindent
expressing $P_B(r, \theta) \equiv P_B(A, R)$ by a change of argument variables.   The coefficients of the two independent gradients in equation (\ref{tsing_1}) must vanish everywhere, giving
\begin{eqnarray}
\label{tsing_2}
\Lambda(A) {\mathcal L} A - v_A^2 {d\phi \over dA} {d^2\phi \over dA^2} |\nabla A|^2 + \frac{1}{2} {d \over d A} \left[ \Lambda(A) S^2_1(A)\right] &&\nonumber \\
-\frac{1}{2}{c^2 R^4 \over 4 \pi \rho_0} {d \over d A}  \left[\Lambda(A) S^2_2(A) \right]  
+ R^2 {\partial \over \partial A} P_B(A, R) = 0 ,&& \\
\label{tsing_3}
2 {c^2 R \over 4 \pi \rho_0} \Lambda(A)S^2_2(A) - {\partial \over \partial R} P_B(A, R) = 0 ,&&
\end{eqnarray}
\noindent
where we have inserted ${\mathcal S}(A, R)$ given by equation (\ref{Q_Psi}).   Integration of Equation (\ref{tsing_3}) with respect to $R$ (holding $A$ constant) yields the Bernoulli pressure
\begin{equation}
\label{Bernoulli2}
P_B(A, R) = P_0 (A) + {c^2 \over 4 \pi \rho_0} R^2 \Lambda(A)S^2_2(A) ,
\end{equation}
\noindent 
where the free generating function $P_0 (A)$ arises as a "constant of integration".  Substituting for $P_B(A, R)$ in Equation (\ref{tsing_2}) then gives the single, second-order elliptic Tsinganos PDE for $A$:
\begin{eqnarray}
\label{tsing_4}
\Lambda(A) {\mathcal L} A - v_A^2 {d\phi \over dA} {d^2\phi \over dA^2} |\nabla A|^2 + \frac{1}{2} {d \over d A} \left[ \Lambda(A) S^2_1(A)\right]&& \nonumber \\+ {c^2  \over 8 \pi \rho_0} R^4 {d \over d A}  \left[\Lambda(A) S^2_2(A) \right] + R^2 {d \over d A} P_0(A) = 0&& ,  
\end{eqnarray}
\noindent
subject to boundary conditions appropriate for a steady star.  The Tsinganos PDE is defined by a prescribed set of free functions, $\left[\phi(A), S_1(A), S_2(A), P_0(A) \right]$, hereafter the PDF free functions.  The flow ${\bf v}$, field ${\bf B}$ and fluid pressure $p$ are given in terms of solution $A$, with $p$ related to Bernoulli pressure $P_B$ by equation (\ref{Bernoulli_P}).   

Consider a field of a fixed topology ${\mathcal T}$ embedded in a star of total mass $M_0$ and uniform density $\rho_0$ that fix the stellar volume $V_0 = M_0/\rho$.  The properties $\left[{\mathcal T}, M_0, \rho_0\right]$ define a physically specific fluid, these properties unchanged in whatever physical state this fluid finds itself with $V_0$ taking a particular shape of its free boundary $\partial V$.  The incidental, steady stellar-flow satisfying PDE (\ref{tsing_4}) for this physically specific fluid is beyond the scope of the study.  The free functions $\left[\phi(A), S_1(A), S_2(A), P_0(A) \right]$ as well as the shape of $\partial V$ are properties of the desired steady flow that cannot be prescribed {\it a priori} because they must be treated as mathematical unknowns, to be solved with a solution of PDE (\ref{tsing_4}) that recovers the given $\left[{\mathcal T}, M_0, \rho_0\right]$.  In fact, the constraints imposed by a given physical fluid is far more stringent.  The frozen-in condition dictates a specific mass conserved between any two flux surfaces, described by a continuum of integral equations that are coupled to PDE (\ref{tsing_4}), to be solved simultaneously for $A$, $\left[\phi(A), S_1(A), S_2(A), P_0(A) \right]$ and the shape of $\partial V$.  

Modest progress can be made by constructing tractable solutions of the Tsinganos PDE for a given fluid domain $V$, defined by a prescription of (i) the shape of $\partial V$ and (ii) the free functions $\left[\phi(A), S_1(A), S_2(A), P_0(A) \right]$ that render the Tsinganos PDE linear and solvable by standard mathematical methods.   The boundary $\partial V$ must be stress-free under conditions (\ref{stress-free_condition1}) to relate the steady flow to the exterior vacuum.  Generally, these stress-free conditions are stringent so that the solution to the free-boundary problem may not exist for a prescribed set of PDF free functions, a nonlinear  property.  Each proper solution to the free-boundary problem constructed defines a physically specific stellar steady flow with its invariant field topology.  We proceed to construct such steady flows in the simplest case of a spherical star of given radius $r_0$, that is, the stress-free boundary $\partial V$ incidentally taking a spherical shape. 

\subsection{A family of steady cross-field flows}

To render Tsinganos PDE (\ref{tsing_4}) linear, we prescribe the poloidal stream-function
\begin{equation}
\label{phi_Alin}
\phi(A) = \gamma_0 A , 
\end{equation}
\noindent
describing the poloidal alignment of ${\bf v}_p$ and ${\bf B}_p$ with $\gamma_0$ a given constant.  By the normalization in use, $v_0 = v_A = B_0 \left( 4 \pi \rho_0 \right)^{-1/2}$,  
\begin{equation}
\label{v_pB_p}
{\bf v}_p = {\gamma_0 \over \left(4 \pi \rho_0 \right)^{1/2}} {\bf B}_p ,
\end{equation}
\noindent
with the factor $\Lambda(A) = \lambda_0$ being a constant in equations (\ref{PsiQ3}):
\begin{equation}
\label{lambda_0}
\lambda_0 = v_A^2 \left( 1 - \gamma_0^2\right) .  
\end{equation}
\noindent
Relative to the poloidal Alfv\`{e}n speed $v_A$, the poloidal flow ${\bf v}_p$ is everywhere sub-Alfv\`{e}nic if $\lambda_0 > 0, \gamma_0^2 < 1$ and super-Alfv\`{e}nic if $\lambda_0 < 0, \gamma_0^2 > 1$; see equation (\ref{supersub-Alfven}), parametrically separated by the special case $\lambda_0 = 0, \gamma_0^2 = 1$ .

Assuming $\lambda_0 \ne 0, \gamma_0^2 \ne 1$, we next prescribe the other generating functions, the $\varphi$-force integral $K(A)$, electrostatic potential $W(A)$, and Bernoulli-pressure profile $P_0(A)$:
\begin{eqnarray}
\label{K_linA}
K(A) &\equiv& \Lambda(A) S_1(A) = \lambda_0 s_1 A , \\
\label{W_1.5A}
W(A) &=&\lambda_0\left( {8 \pi \rho_0 \over c^2} \right)^{1/2}{2 \over 3} s_2 A |A|^{1/2} , \\
\label{P_1lin}
P_0(A) &=& p_0 + \lambda_0 p_1 A , 
\end{eqnarray}
\noindent
where $s_1, s_2, p_0, p_1$ are positive constants.  The electrostatic potential $W(A)$ is a continuous, monotonically-increasing odd function of $A$ with the following properties:
\begin{eqnarray}
\label{dW_dA}
{dW(A) \over dA} &\equiv& \Lambda(A) S_2(A) \nonumber \\
&=& \lambda_0 \left( {8 \pi \rho_0 \over c^2} \right)^{1/2} s_2 |A|^{1/2} , \\
\label{d2W_dA2}
{d^2W \over dA^2} &=& \lambda_0{\mathcal H}(A)\left( {8 \pi \rho_0 \over c^2} \right)^{1/2}\frac{s_2}{2} |A|^{-1/2} ,  \\
\label{grad_dWdAsq}
{d \over dA} \left( {dW \over dA} \right)^2 &=& {\mathcal H}(A) \left( {8 \pi \rho_0 \over c^2} \right) s_2^2, \\\label{Heaviside}
{\mathcal H}(A) &=& \pm 1 ~~~ \mathrm{if} ~A > 0 ~\mathrm{or} ~ A < 0 .
\end{eqnarray}
\noindent
The first derivative of $W(A)$ is a continuous even function of $A$, vanishing at $A = 0$ and positive elsewhere whereas the second derivative is an odd function of $A$, discontinuous and unbounded at $A = 0$, where ${\mathcal H}(A)$ is the Heaviside function. 

With $S_1(A) = s_1 A$ and $S_2(A) = \left( {8 \pi \rho_0 \over c^2} \right)^{1/2} s_2 |A|^{1/2}$, 
equations (\ref{PsiQ6}) and (\ref{PsiQ7}) defining $v_{\varphi} = r \sin \theta~\Psi$ and $B_{\varphi} = r \sin \theta~Q$ give 
\begin{eqnarray}
\label{PsiQ8}
\Psi &=&  \gamma_0 s_1 A - r^2 \sin^2 \theta s_2 \sqrt{2|A|} , \\ 
\label{PsiQ9}
Q &=& s_1 A - \gamma_0 r^2 \sin^2 \theta s_2 \sqrt{2|A|} ,
\end{eqnarray}
\noindent
and PDE (\ref{tsing_4}) gives the governing PDE for $A$:
\begin{equation}
\label{tsing_5}
\lambda_0 \left[ {\mathcal L} A  + s_1^2 A + {\mathcal H}(A) s^2_2 r^4 \sin^4 \theta + p_1 r^2 \sin^2 \theta \right] = 0 , 
\end{equation}
\noindent 
accounting for the discontinuity of ${d2W \over dA^2}$ across $A = 0$ described by equations (\ref{dW_dA})-(\ref{grad_dWdAsq}).  Factoring away $\lambda_0 \ne 0$, an assumption, the governing PDE of $A$ is then
\begin{equation}
\label{tsing_6}
{\mathcal L} A  + s_1^2 A + {\mathcal H}(A) s^2_2 r^4 \sin^4 \theta + p_1 r^2 \sin^2 \theta = 0 .
\end{equation}
\noindent 
The constants $p_0$ and $p_1$ define the Bernoulli pressure given by equations (\ref{Bernoulli2}) and (\ref{P_1lin}):  
\begin{equation}
\label{Bernoulli3}
P_B(A, R) = p_0 + \lambda_0 \left( p_1 + 2 \lambda_0 s_2^2 R^2 \right) A ,
\end{equation}
\noindent 
in terms of which the fluid pressure $p$ is given by equation (\ref{Bernoulli_P}).

Although the two branches of PDE (\ref{tsing_6}) are linear in $A$, they pose a nonlinear, global free-boundary problem that requires solving the two branch-PDEs simultaneously.  Which PDE to apply depends on where in the stellar volume $V$ the mathematically unknown global solution $A$ takes one of the two algebraic signs.  This nonlinear problem is formidably beyond the analytical reach of the study but understanding its mathematical structure in broad terms is conceptually important.  If the global solution $A$ takes opposite algebraic signs in $V$, the solution field is composed of multiple flux-systems, each bounded by a stress-free flux-surfaces $A = 0$ and described by $A$ of a particular sign governed by the relevant branch PDE ascribed by the Heaviside function ${\mathcal H}(A)$.  In general terms, each flux system is governed by Tsinganos PDEs (\ref{tsing_2}) and (\ref{tsing_3}) defined by a prescription of the free functions $\left[\phi(A), S_1(A), S_2(A), P_0(A) \right]$, with all the inter-system interior boundaries $A = 0$ as well as the $A = 0$ stellar boundary $\partial V$ subject to the stress-free conditions.  Thus, the formidable problem is to construct a juxtaposition of steady-flow flux systems that are contiguous across stress-free flux-surfaces $A = 0$.  We limit our study to an instructive simpler problem, tentatively solving PDE  (\ref{tsing_6}) for $A$ in an entire stellar volume $V$ with a given spherical shape of the stress-free boundary $\partial V$.

\subsection{Tsinganos-Ferraro cross-field steady flows}

To avoid the complication of the shape of $\partial V$ being a mathematical unknown, let us solve PDE (\ref{tsing_6}) for a family of steady cross-field flows, hereafter the Tsinganos-Ferraro (TsF) flows, in the case of a star that is incidentally spherical with radius $r_0$ with no field extending into the $r > r_0$ vacuum.  The stress-free boundary conditions (\ref{stress-free_condition1}) give
\begin{equation} 
\label{general_bc2}
r = 0, ~~A = 0; ~~~~r = r_0, ~~A = {\partial A \over \partial r} = 0 ,
\end{equation}
\noindent
identifying $\partial V$ as a spherically-shaped flux-surface $A = 0$ that is stress-free by virtue of the normal derivative of $A$ vanishing on $\partial V$.  

PDE (\ref{tsing_6}) may be rewritten as 
\begin{eqnarray}
\label{tsing_7}
&& {\mathcal L} A  +s_1^2 A - \frac{1}{5}s^2_2 r^4 {\mathcal P}_3 (\mu) +  \left\{ \frac{4}{5}s^2_2 r^4 +  p_1 r^2 \right\} {\mathcal P}_1 (\mu) = 0 , \nonumber \\
\\
\label{P1_mu}
&& {\mathcal P}_1 (\mu) = \left(1 - \mu^2 \right)^{1/2} P_1^1(\mu) = \left(1 - \mu^2 \right), \\
\label{P3_mu}
&& {\mathcal P}_3 (\mu) =  \left(1 - \mu^2 \right)^{1/2} P_3^1(\mu) = \left(1 - \mu^2 \right)\left(5 \mu^2 -1 \right),
\end{eqnarray}
\noindent 
in terms of the $n=1$ and $n=3$ associated Legendre functions $P_n^1(\mu)$, $\mu = \cos \theta$.  The functions ${\mathcal P}_1 (\mu)$ and ${\mathcal P}_3 (\mu)$ being orthogonal over the interval $-1 < \mu < 1$,  solution $A$ may be expressed as a linear superposition 
\begin{eqnarray}
\label{A_A1A3}
A &=& A_1 + A_3 \nonumber \\ 
&\equiv& F_1(\zeta) {\mathcal P}_1 (\mu) + F_3(\zeta) {\mathcal P}_3 (\mu) , 
\end{eqnarray}
\noindent  
in terms of the variables
\begin{equation}
\label{zeta}
\zeta = s_1 r , ~~~\zeta_0 =  s_1 r_0 . 
\end{equation}
\noindent
PDE (\ref{tsing_7}) thus decomposes into the two inhomogeneous linear ODEs:
\begin{eqnarray}
\label{ODE1}
{d^2 F_1 \over d\zeta^2} + \left( 1 - { 2 \over \zeta^2}\right) F_1 +  {4 s_2^2 \over 5 s_1^6} \zeta^4 + {p_1 \over s_1^2} \zeta^2 = 0 , \\
\label{ODE3}
{d^2 F_3 \over d\zeta^2} + \left(1 - {12 \over \zeta^2} \right) F_3 -  {s_2^2 \over 5 s_1^6} \zeta^4 = 0 , 
\end{eqnarray}
\noindent  
subject to the three sets of homogeneous boundary-conditions:
\begin{eqnarray}
\label{ODE_bc1}
&&\mathrm{at} ~~\zeta = 0, ~~~ F_1 = F_3 = 0 , \\
\label{ODE_bc2}
&&\mathrm{at} ~~\zeta = \zeta_0, ~~~ F_1 = F_3 = 0 , \\
\label{ODE_bc3}
&&\mathrm{at} ~~\zeta = \zeta_0, ~~~ {dF_1 \over d\zeta} = {dF_3 \over d\zeta} = 0 , 
\end{eqnarray}
\noindent  
under stress-free conditions (\ref{general_bc2}).

Consider the homogeneous ODEs
\begin{equation}
{d^2 f_n \over d\zeta^2} + \left(1 - {n(n+1) \over \zeta^2}\right) f_n = 0 ,
\end{equation}
\noindent
$n = 1, 2, 3, ...$, with a pair of independent solutions $f_{\pm n} = \zeta j_{\pm n}(\zeta)$, expressed in terms of spherical Bessel functions.  ODEs (\ref{ODE1}) and (\ref{ODE3}) have the general solutions
\begin{eqnarray}
\label{F1_gen}
F_1(\zeta) &=& a_{-1} \zeta j_{-1} (\zeta) + a_1 \zeta j_1(\zeta) - \frac{4}{5} {s_2^2 \over s_1^6} \zeta^4 + \left[ 8 {s_2^2 \over s_1^6} - {p_1 \over s_1^2} \right] \zeta^2 \nonumber \\
\label{F3_gen}
F_3(\zeta) &=&  a_{-3}\zeta j_{-3}(\zeta) + a_3\zeta  j_3(\zeta) + \frac{1}{5}{s_2^2 \over s_1^6}\zeta^4 , 
\end{eqnarray}
\noindent 
where $a_{\pm 1}$ and $a_{\pm 3}$ are constant amplitudes.  We set $a_{-1} = a_{-3} = 0$ under boundary conditions (\ref{ODE_bc1}).  The remaining 5 free constants $a_1, a_3, s_1, s_2, p_1$ are available for meeting the other 2 pairs of boundary conditions (\ref{ODE_bc2}) and (\ref{ODE_bc3}), which determine 
\begin{eqnarray}
\label{a_1}
a_1 j_2(\zeta_0) &=&  - \frac{8}{5}{s_2^2 \over s_1^6} \zeta_0^2 , \\
\label{a_3}
a_3 &=& -\frac{1}{5}{s_2^2 \over s_1^6} {\zeta_0^3 \over j_3(\zeta_0)} , \\ 
\label{p_1}
p_1 &=& \frac{4}{5}{s_2^2 \over s_1^4}\left[10 - \zeta_0^2 - 2 \zeta_0 {j_1(\zeta_0) \over j_2(\zeta_0)} \right] ,\end{eqnarray}
\noindent  
giving the solutions and their first derivatives:
\begin{eqnarray}
\label{F1}
F_1(\zeta) &=& \frac{4}{5} {s_2^2 \over s_1^6}\left[ \zeta^2 (\zeta_0^2 - \zeta^2) +2 \zeta_0 \zeta {\zeta j_1(\zeta_0) - \zeta_0 j_1(\zeta) \over j_2(\zeta_0)} \right] ,\nonumber \\
\\
\label{dF1}
{dF_1(\zeta) \over d\zeta} &=& \frac{4}{5}{s_2^2 \over s_1^6} \left[ 2 \zeta_0^2 \zeta - 4 \zeta^3 \right.\nonumber \\
&&~~~~~~\left. + 2 \zeta_0 {2 \zeta j_1(\zeta_0) - 2 \zeta_0 j_1(\zeta) + \zeta_0 \zeta  j_2(\zeta) \over j_2(\zeta_0)} \right] , \nonumber \\
\\
\label{F3}
F_3(\zeta) &=& -\frac{1}{5}{s_2^2 \over s_1^6} \left[ \zeta_0^3 \zeta {j_3(\zeta) \over j_3(\zeta_0)} - \zeta^4 \right]  , \\
\label{dF3}
{dF_3(\zeta) \over d\zeta} &=&  -\frac{1}{5}{s_2^2 \over s_1^6} \left[ 4\zeta_0^3{j_3(\zeta) \over j_3(\zeta_0)}  - 4 \zeta^3 - j_4(\zeta) \right] . 
\end{eqnarray}
\noindent 
Under boundary conditions (\ref{ODE_bc3}), equation (\ref{dF3}) imposes the condition
\begin{equation}
\label{j4_bc}
\mathrm{at} ~~\zeta = \zeta_0, ~~~ {dF_3 \over d\zeta} =  \frac{1}{5}{s_2^2 \over s_1^6}j_4(\zeta_0) = 0 .
\end{equation} 
Therefore, given radius $r_0$, the free parameter $s_1$ defining $\zeta_0 = s_1 r_0$ cannot be arbitrarily prescribed. Nevertheless, there are infinitely many solutions to the boundary-value problem, generated by the eigenvalue equation 
\begin{equation}
\label{j4_eigenvalues}
j_4(\zeta_{4,m}) = 0 , ~~~ m = 1, 2, 3, ... ,
\end{equation}  
\noindent
defining $\zeta_0 = s_1 r_0 = \zeta_{4,m}$ as the eigenvalue for eigenfunction $A =  A^{(m)}(r, \theta) = A_1^{(m)}(r, \theta) + A_3^{(m)}(r, \theta)$ given by equation (\ref{A_A1A3}), setting
\begin{eqnarray}
\label{A1,m}
A_1 &=& A_1^{(m)}(r, \theta) \nonumber \\ 
&=& {s_2^2 \over s_1^6}\frac{4}{5}\left[ \zeta^2 (\zeta_{4,m}^2 - \zeta^2) \right. \nonumber \\
&&\left.~~~+ 2 \zeta_{4,m} \zeta {\zeta j_1(\zeta_{4,m}) - \zeta_{4,m} j_1(\zeta) \over j_2(\zeta_{4,m})} \right] \sin^2 \theta , \\
\label{A3,m}
A_3 &=& A_3^{(m)}(r, \theta) \nonumber \\
&=& {s_2^2 \over s_1^6} \frac{1}{5}\left[ \zeta^4 - \zeta_{4,m}^3 \zeta {j_3(\zeta) \over j_3(\zeta_{4,m})}\right] \sin^2 \theta \left( 4 - 5 \sin^2 \theta \right) . \nonumber \\
\end{eqnarray}
Each eigenfunction $A = A^{m}(r, \theta)$ describes a poloidal field ${\bf B}_p$ in a TsF flow by the normalization $B_0 = {s_2^2 \over s_1^6}$, which defines a reference poloidal Alfv\`{e}n speed $v_A = B_0\left(4 \pi \rho_0 \right)^{-1/2}$.  The parameter $s_1$ is defined by the eigenvalue $\zeta_0 = s_1 r_0 = \zeta_{4,m}$ so that prescribing the amplitude $B_0$ of the ${\bf B}_p$ fixes the parameter $s_2$.  The poloidal velocity ${\bf v}_p = {\gamma_0 \over \sqrt{4 \pi \rho_0}} {\bf B}_p$ ; see equation (\ref{v_pB_p}), is then fixed by the single the free parameter $\gamma_0^2 \ne 1$, with $\lambda_0 = v_A^2 (1 - \gamma_0^2) \ne 0$.  To complete the description of the TsF $\left({\bf v}, {\bf B}\right)$-flow, the parameters $\left(s_2, \gamma_0\right)$ define the azimuthal components given by $\Psi = r \sin \theta v_{\varphi}$ and $Q = = r \sin \theta B_{\varphi}$ where 
\begin{eqnarray}
\label{PsiQ10}
\Psi &=&  \gamma_0 s_1 A^{(m)} - r^2 \sin^2 \theta s_2 \sqrt{2|A^{(m)}|} , \\ 
\label{PsiQ11}
Q &=& s_1 A^{(m)} - \gamma_0 r^2 \sin^2 \theta s_2 \sqrt{2|A^{(m)}|} ;
\end{eqnarray}
\noindent
see equations (\ref{PsiQ8}) and (\ref{PsiQ9}).

\begin{figure*}
\label{Fig1}
\centerline{\includegraphics[width=90mm]{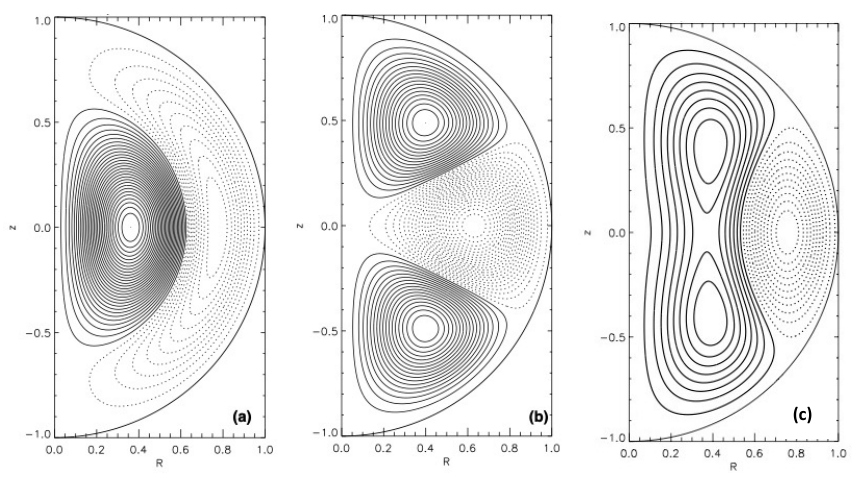}}
\caption{\small{Contour plots of component flux-functions (a) $A_1^{(m=1)}(r, \theta)$, (b) $A_3^{(m=1)}(r, \theta)$ and their superposition giving the physical poloidal flux function (c) $A = A^{(m=1)}(r, \theta)$, generated by the first eigenvalue $s_1 r_0 = \zeta_0 = \zeta_{4,(m=1)} = 8.1826$, employing cylindrical coordinates $(R, z) \equiv (r \sin \theta, r \cos \theta)$ and setting $r_0 = 1$.  All three flux functions and their radial derivatives vanish at the stress-free boundary $\partial V: r = r_0$.  Solid (dot) contours indicate positive (negative) flux-functions.  The closed contours of $A$ in (c) are the field lines of poloidal ${\bf B}_p$ of constant amplitude $B_0$ in the $m = 1$ TsF flow, circulating clockwise or anti-clockwise for positive or negative $A$, respectively.  The poloidal ${\bf v}_p$ in the TsF flow is aligned with ${\bf B}_p$ with proportionality $\gamma_0$, described in the text.  In 3D space, the closed $A$-contours sweep out the toroidal surfaces of axisymmetric flux-tubes, the flux function $A$, stream function $\phi (A)$ and the electric potential $W(A))$ taking constant values on each flux-tube surface. Described in the text, {\it most} of the mutually-crossing field and flow lines on each constant-$A$ flux-tube surface are infinitely long, winding endlessly around the flux-tube axis and around the axis of symmetry with the azimuthal components $B_{\varphi} \ne v_{\varphi}$ given by equations (\ref{PsiQ10}) and (\ref{PsiQ11}).}\\ \\ \\}
\end{figure*}

\begin{figure*}
\label{Fig2}
\centerline{\includegraphics[width=90mm]{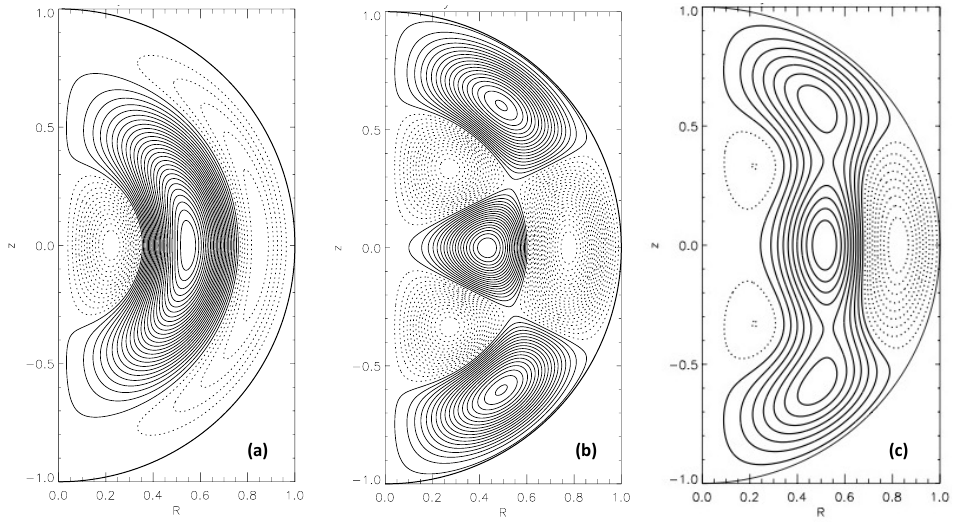}}
\caption{\small{Contour plots of component flux-functions (a) $A_1^{(m=2)}(r, \theta)$, (b) $A_3^{(m=2)}(r, \theta)$ and their superposition giving the poloidal flux function (c) $A = A^{(m=2)}(r, \theta)$, generated by the second eigenvalue $s_1 r_0 = \zeta_0 = \zeta_{4,(m=2)} = 11.7049$, employing cylindrical coordinates $(R, z) \equiv (r \sin \theta, r \cos \theta)$ and setting $r_0 = 1$; presented in the same format and description as in Fig. 1.}\\ \\ \\}
\end{figure*}
 
Respectively displayed in Figs. 1 and 2 are the physical flux-functions $A = A^{(m)}(r, \theta)$ for the first and second eigenvalues $\zeta_{4,m = 1} = 8.1826$ and $\zeta_{4,m = 2} = 11.7049$.  Shown in each figure are the pair of component flux-functions, $A_1 = A_1^{(m)}(r, \theta)$ and $A_3 = A_3^{(m)}(r, \theta)$ whose linear superposition gives $A = A^{(m)}(r, \theta)$, the solid and dot contours respectively denoting positive and negative values.  These contours describe closed poloidal field-lines circulating clockwise and anti-clockwise for positive and negative flux function, respectively.  

In Fig. 1, the poloidal field $A = A^{(m=1)}(r, \theta)$ comprises two nests of closed field lines, an $A < 0$ outer single-nest sandwiched between the boundary $r = r_0$ and an $A > 0$ inner double-nest.  In Fig. 2, the poloidal field $A = A^{(m=2)}(r, \theta)$ has three nests of closed field lines, an $A > 0$ triple-nest sandwiched between an $A < 0$ outer single-nest and an $A <  0$ inner double-nest.  Each flux function $A = A^{(m)}(r, \theta)$ by its interior flux surfaces $A = 0$ partitions the stellar volume $V$ into partial volumes $V_{\pm}$ where $A > 0$ or $A < 0$, respectively, hereafter referred to as the positive and negative sub-volumes. 

We address a complication, that flux functions $A = A^{(m)}(r, \theta)$ is not a complete global solution because it is calculated globally from the $A > 0$ branch of PDE (\ref{tsing_6}).  The solution $A = A^{(m)}(r, \theta)$ is thus meaningful in the $V_{(+)}$ sub-volumes, requiring the $V_{(-)}$ sub-volumes to be occupied by a suitable re-defined steady-flows with $A < 0$.  

We tentatively proceeded by solving PDE (\ref{tsing_7}) over the entire stellar volume $V$.  Had the global solution thus obtained turned out to be positive everywhere in $V$, there would be no issue then.  But, this is not the case, so that the steady solutions $A = A^{(m)}(r, \theta)$ in the $V_{(-)}$ sub-volumes must be rejected and replaced with suitably re-constructed steady-flow solutions.   Inserting steady-flow solutions of the $A < 0$ branch of PDE (\ref{tsing_6}) that geometrically fill the $V_{(-)}$ sub-volumes is possible but the stress-free condition is unlikely to be satisfied at the $A = 0$ boundary separating the $V_{\mp}$ sub-volumes.  Accepting that simultaneously solving the two branches of PDE (\ref{tsing_6}) is not tractable, we limit the purpose of the study to investigate the $A = A^{(m)}(r, \theta)$ solution in the $V_{(+)}$, taking the $V_{(-)}$ sub-volumes to conceptually represent an (unspecified) complementary parts of a global flow.  

Gravity provides the sole means of stellar confinement against breakup, both the field and flow being globally expansive.  Integrating Newton's equation (\ref{Newton}) gives the the gravitational potential of the spherical star
\begin{eqnarray}
\label{gravitationalU}
U (r) &=& - {G M_0\over r_0} +\frac{1}{2} {G M_0\over r_0^3} \left(r^2 -  r_0^2 \right)~ ~\mathrm{in} ~ r \le r_0 , \nonumber \\ 
&=& -{G M_0 \over r} ~~\mathrm{in} ~ r > r_0 .
\end{eqnarray}
\noindent 
The Bernoulli pressure $P_B$ defined by equations (\ref{Bernoulli_P}), (\ref{Bernoulli2}) and (\ref{dW_dA}) gives the pressure:
\begin{equation}
\label{pressure}
p =  \rho_0 \left\{p_0 + \lambda_0 \left(p_1 + 2 s_2^2 R^2 \right) A - U(r) - \frac{1}{2} v^2 \right\} , ~~~ r \le r_0 ,
\end{equation}
\noindent
the parameter $p_1$ given by equation (\ref{p_1}).  In Tsinganos PDE (\ref{tsing_6}), the parameter $p_1$ cannot be freely prescribed to define this PDE, its value to be fixed by the solution $A$ of the free-boundary problem, underpinning the nonlinear inter-dependence between the variables and free-boundary conditions.  

Imposing the vacuum condition $p = 0$ at $r = r_0$, where $A = 0$, $v = 0$, and $U(r_0) = - {G M_0 \over r_0}$, the constant parameter
\begin{equation}
\label{P_0_condition}
p_0 = -{G M_0 \over r_0} ,
\end{equation}
\noindent
is fixed to express the pressure as 
\begin{equation}
\label{P_0_condition2}
p + \frac{1}{2} \rho_0 v^2 = \rho_0 \left\{ \frac{1}{2} {G M_0\over r_0^3} \left(r_0^2 -  r^2 \right) + \lambda_0 \left( p_1 + 2 s_2^2 R^2 \right) A \right\}, ~~~ r \le r_0  .
\end{equation}
\noindent
In the absence of flow and field, we have the spherically-symmetric hydrostatic pressure 
\begin{equation}
\label{P_0_condition3}
p = \frac{1}{2} {G M_0 \rho_0 \over r_0^3} \left(r_0^2 -  r^2 \right), ~~~ r \le r_0 ,
\end{equation}
\noindent
that monotonically and radially decreases from a maximum at $r = 0$ to zero at $r = r_0$, its outward force supporting the self-gravitational weight of the fluid.  In the presence of a TsF flow, the star is held together by self-gravity with its otherwise spherically symmetric pressure locally reduced so as to wholly confine the stellar flow/field, expressed by equation (\ref{P_0_condition2}).  The pressure $p$ physically cannot be negative.  Therefore,  equation (\ref{P_0_condition2}) sets calculable bounds, not needed in our discussion here, on the flow and field to meet the condition $p \ge 0$ in $r < r_0$ for a given $r_0$ and total mass $M_0 = \frac{4}{3} \rho_0 r_0^3$.  Clearly, these bounds are met for sufficiently weak flows and fields. 

Several points are noteworthy, relating the particular properties encountered in the TsF steady flows to general free-boundary problems.  The spherical shape of the TsF star is incidental, postulated to simplify the mathematical problem.  When perturbed, the free-boundary $\partial V$ is readily deformed out of its incidental spherical shape.  As a separate point, the steady spherical TsF construction may be generalized to oblate and prolate spheroidal shapes of the boundary $\partial V$.  

Subject to the juxtaposition of TsF cross-field flows in $V_{(+)}$ with a suitably constructed steady-flow in the sub-volume $V_{(-)}$, the global steady flow so constructed has a specific invariant field-topology ${\mathcal T}$.   Perturbing such a steady star out of its incidentally-spherical  boundary shape under the frozen-in condition, does not change the topology ${\mathcal T}$ provided the field remains wholly contained.  

\subsection{The TsF $m = 1$ cross-field flows}

Four TsF flows described by the $m = 1$ eigenfunction $A = A^{(m=1)}(r, \theta)$ with eigenvalue $s_1 r_0 = \zeta_{4,m=1} = 8.1826$, are presented in Figs. 3 and 4 in terms of  the contours of azimuthal components $\Psi = r \sin \theta  v_{\varphi}$ and $Q = r \sin \theta B_{\varphi}$ for parameter $\gamma_0 = \mp 0.5, \mp 1.5$.  The properties are plotted for the entire mathematical solution $A^{(m=1)}(r, \theta)$, with the two sub-volumes $V_{(\pm)}$ separated by the poloidal field-line $A^{(m=1)}(r, \theta) = 0$.  We keep in mind that the eigenfunction describes a physical TsF steady flow in $V_{(+)}$ where $A^{(m=1)}(r, \theta) >  0$.  Whereas, the solution $A^{(m=1)}(r, \theta) < 0$ in $V_{(-)}$ is a nominal representation of a steady flow, intractable to construct as explained in the text.  We henceforth focus attention on the physical flow in the sub-volume $V_{(+)}$.

In Figs. 3 and 4, a pair of (thick) contours of $A = A^{(m=1)}(r, \theta)$ are inserted in sub-volume $V_{(+)}$, indicated by arrows, to aid visualizing the variations of $Q$ and $\Psi$ along a constant-$A$ poloidal field line.  The poloidal ${\bf B}_p$ and ${\bf v}_p$ are respectively parallel or anti-parallel for positive and negative $\gamma_0$, but ${\bf B}$ and ${\bf v}$ are not aligned, with unequal azimuthal components, i.e., $B_{\varphi} \ne v_{\varphi}$.  

For $\gamma_0 < 0$, $\left(B_{\varphi}, v_{\varphi}\right)$ are each uniformly of the same algebraic sign in $V_{(+)}$.  In this case, each of the 3D field and flow lines of ${\bf B}$ and ${\bf v}$ winds around the axis of symmetry progressively without turning back.  For $\gamma_0 > 0$, $\left(B_{\varphi}, v_{\varphi}\right)$ are each not uniformly of the same algebraic sign in $V_{(+)}$.  In this case, each of the the 3D field and flow lines of ${\bf B}$ and ${\bf v}$ winds progressively around the axis of symmetry, say, froward in the $\varphi$-direction, only to reverse direction and then to reverse direction again to continue on in the forward $\varphi$-direction.  
 
\begin{figure*}
\label{Fig3}
%\centerline{\includegraphics[width=150mm]{L&M1gmp0.5.pdf}}
\centerline{\includegraphics[width=120mm]{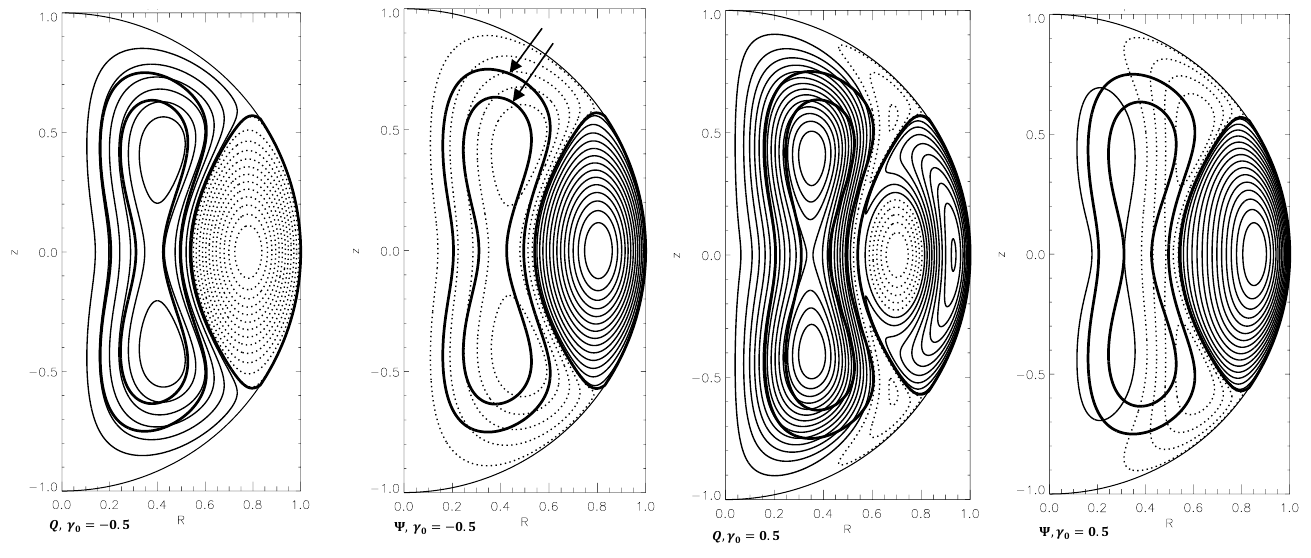}}
\caption{\small{Contour-plots of $Q = r \sin \theta B_{\varphi}$ and  $\Psi = r \sin \theta v_{\varphi}$ in the first-eigenvalue $\zeta_0 = \zeta_{4, (m=1)} = 8.1826$ steady-flows located in the $V_{(+)}$ sub-volume where $A^{(m=1)}(r, \theta) > 0$, defined by $\gamma_0 = \mp 0.5$ that respectively set ${\bf v}_p$ anti-parallel and parallel with ${\bf B}_p$.  Solid and dot contours respectively indicate azimuthal components pointing in positive and negative $\varphi$-directions, the $\left( Q, \Psi \right)$ contours to be viewed as profiles along the individual poloidal constant-$A$ field-lines of ${\bf B}_p$ displayed in Fig.1 (c).  The pair of arrows indicate two selected constant-$A$ field lines, as thick contours, inserted in each of the four subfigures to provide a visual reference for the variations of $Q$ and $\Psi$ along a constant-$A$ field line.  The $\gamma_0 = \mp 0.5$ flows shown are sub-Alfv\`{e}nic, with $\lambda_0 = v_A^2 (1 - \gamma_0^2) > 0$.  As explained in the text, the mathematical solution $A^{(m=1)}(r, \theta)$ describes a physical TsF cross-field steady flow in sub-volume $V_{(+)}$ whereas the mathematical solution $A^{(m=1)}(r, \theta) < 0$ in sub-volume $V_{(-)}$ is a nominal representation of a flow required for balancing the forces between the two sub-volumes.}\\ \\ \\}
\end{figure*}
 
\begin{figure*}
\label{Fig4}
%\centerline{\includegraphics[width=150mm]{L&M1g1.5ccQQ01.5&-.5 copy.pdf}}
\centerline{\includegraphics[width=120mm]{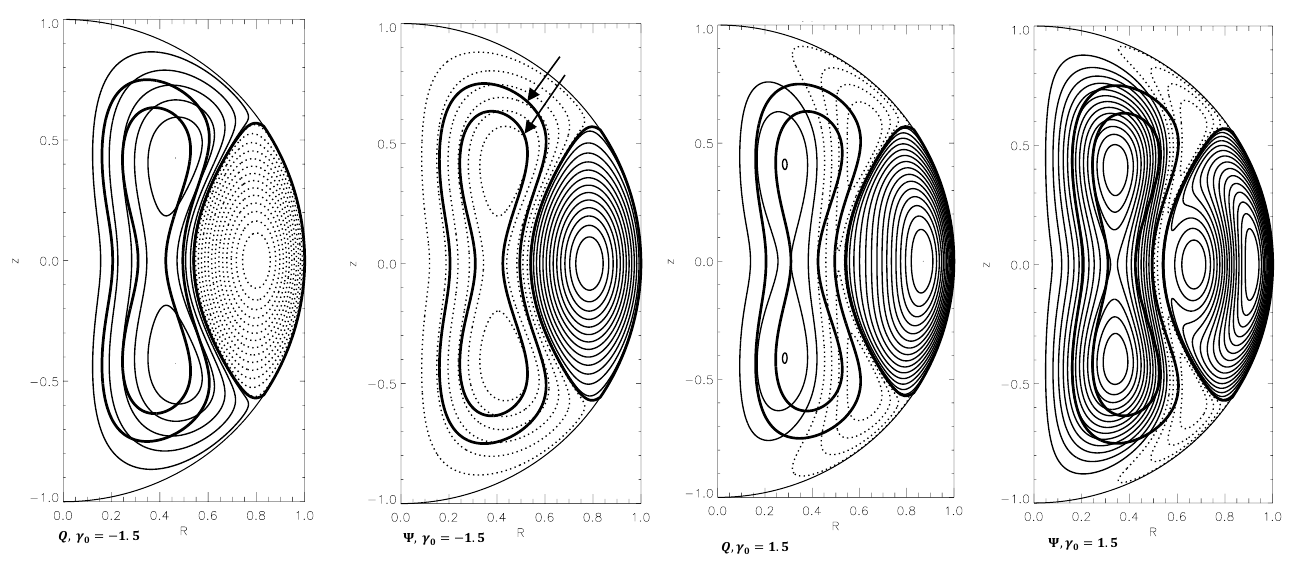}}
\caption{\small{Contour-plots of $Q = r \sin \theta B_{\varphi}$ and  $\Psi = r \sin \theta v_{\varphi}$ in the first-eigenvalue $\zeta_0 = \zeta_{4, (m=1)} = 8.1826$ steady-flows located in the $V_{(+)}$ sub-volume where $A^{(m=1)}(r, \theta) > 0$, defined by $\gamma_0 = \mp 1.5$ that respectively set ${\bf v}_p$ anti-parallel and parallel with ${\bf B}_p$.; displayed in the same format as in Fig. 3.  Both flows are super-Alfv\`{e}nic, with $\lambda_0 = v_A^2 (1 - \gamma_0^2) < 0$.}\\ \\ \\}
\end{figure*}
 
For $\gamma_0^2 < 1, \lambda_0 =  v_A \left(1 - \gamma_0^2\right) > 0$, the flow is sub-Alfv\`{e}nic, whereas for $\gamma_0^2 > 1, \lambda_0 =  v_A \left(1 - \gamma_0^2\right) < 0$, the flow is super-Alfv\`{e}nic.  In spite of their similar flow and field morphologies, the sub-Alfv\`{e}nic and super-Alfv\`{e}nic TsF flows respond to perturbations differently.   In a sub-Alfv\`{e}nic flow, Alfv\`{e}n waves can overtake the fluid flow to spread a perturbation to magnetically connected parts of the fluid.  In a super-Alfv\`{e}nic flow, the Alfv\`{e}n waves cannot overtake a fluid flow.  In the full freedom of 3D space, Alfv\`{e}n waves can accumulate nonlinearly to trap extreme field-line deformations in fluid parcels resulting in TDs as hyperbolic characteristics merge or cross to form mathematical singularities.   In both cases of sub-Alfv\`{e}nic or super-sub-Alfv\`{e}nic TsF flows, the perturbed field can naturally break across $\partial V$ to extend into the exterior vacuum.  We remind ourselves that the unperturbed steady magnetic field being wholly contained in the fluid is an assumption made to keep the steady-flow problem simple.  

\subsection{Mutually-intersecting ergodic flow and field lines}

The field/flow lines of the axisymmetric TsF steady flows are almost all of infinite lengths.   Each of the mutually-crossing field and flow lines on a flux-surface $A = A^*$, a constant, winds around the axis of their toroidal flux-surface and encircles the symmetry axis.  There are two topological possibilities, either a field or flow line has a finite length and returns to a starting point on the flux surface $A = A^*$, or else the line is endless and ergodic, going around the symmetry axis and never returning to {\it any} starting point, whatever the number of turns of the line around the axis of the toroidal flux surface.   

Field-line ODEs (\ref{FLeqns}) describe the $\varphi(\ell)$-displacement of a point along a field line on surface $A = A^*$ as a function of path length $\ell$ defined by $d\ell^2 = dr^2 + r^2 d\theta^2 + r^2 \sin^2 \theta d\varphi^2$, giving 
\begin{eqnarray}
\label{ergodic_FLeqns}
\varphi(\ell) &=& {1 \over 2 \pi} \int_0^{\ell} \left[ {B_{\varphi} \over |{\bf B}|} \right]_{A = A^*} {d\ell \over r \sin \theta}\nonumber \\
&=& {1 \over 2 \pi} \int_0^{\ell} \left[ {Q \over \sqrt{|\nabla A|^2 + Q^2}} \right]_{A = A^*} {d\ell \over r \sin \theta} ,
\end{eqnarray}
\noindent
starting from a point on surface $A = A^*$ designated to be $\ell = 0$ located at an arbitrarily chosen $\varphi$-plane designated as $\varphi = 0$, and measuring $\varphi(\ell)$ in unit of $2 \pi$-radian angular displacement.  Suppose the field line is closed with a finite total length $\ell_0 < \infty$.  Then, the line is characterized by the following two path lengths.  The path length $\ell_1$ takes field line once around the symmetry axis,
\begin{equation}
\label{ell_1}
{1 \over 2 \pi} \int_0^{\ell_1} ~ \left[ {Q \over \sqrt{|\nabla A|^2 + Q^2}} \right]_{A = A^*} {d\ell \over r \sin \theta} = 1 .
\end{equation}
\noindent
The path length $\ell_2$ takes the field line once around the toroidal axis of surface $A = A^*$, effecting a $\varphi$-displacement, 
\begin{equation}
\label{ell_2_delta_varphi}
\Delta \varphi (\ell_2) =  {1 \over 2 \pi} \int_0^{\ell_2} ~ \left[ {Q \over \sqrt{|\nabla A|^2 + Q^2}} \right]_{A = A^*} {d\ell \over r \sin \theta} . 
\end{equation}
\noindent
For the finite-length field line to return to any chosen starting point on flux surface $A = A^*$, the path length $\ell_0$ must account for an integer $N_1$ full-turns around the symmetry axis and integer $N_2$ full-turns around the toroidal axis of the flux surface.  That is, $\ell_0 = N_1 \ell_1 = N_2 \ell_2$ which implies that the aspect ratio $\Re = {\ell1 \over \ell_2} = {N_2 \over N_1}$ must be a rational fraction, a global property generally not met for a given field ${\bf B}$.  Closed field-lines do exist but only in isolated flux surface $A = A^*$ of a given field with ergodic endless field-lines on the infinitesimally adjacent flux surfaces. 

The field lines of an axisymmetric field are global geometric objects, each individual field line not necessarily axisymmetric. A closed field line with a rational aspect ratio $\Re > 1$ is obviously not axisymmetric, making more than one turn around the toroidal axis for each circulation around the symmetry axis.  A single ergodic field line densely fills up its entire flux surface, without returning to any starting point however many times it goes around the symmetry axis.  The field observed along the endless field line is ${\bf B}(r, \theta)$, meaning that along the line, the field varies with $(r, \theta)$ but is independent of spatial coordinate $\varphi$.  

Two topological notes, the first is the geometric fact that a line is locally always the intersection between two surfaces.  For field lines, this geometric relation generally is true only as a local property, the basis for the local field representation (\ref{Euler_rep}) in terms of Euler potentials.  The TsF flow by it axis-symmetry has only one set of global flux surfaces defined by $A(r, \theta)$.  Every field line, including the ergodic field lines, is nevertheless an intersection between one of these global flux surfaces and a locally-defined flux surface. The latter flux surface cannot be global, to avoid the absurdity of this single surface intersecting everywhere on the entire constant-$A(r, \theta)$ flux surface \citep{Rosner1989}.  

The other note is that a TD forming in a perturbed TsF steady flow would, by resistive dissipation and field reconnection owing to a weak resistivity, would be reconnecting three-dimensional ergodic field-lines, producing 3D field lines that are ergodic in sub-volumes of space.  The flux surfaces of constant $A$ well defined in the unperturbed TsF steady flow would then cease to be global surfaces as the result of reconnection.   A single, volumetrically-ergodic field line so created would densely fill a sub-volume of space, the single line along its infinite length getting as infinitesimally close as desired to any point in the sub-volume\citep{dombre1986}.  

A similar construction applying ODEs (\ref{FLeqns}) to the incompressible flow gives $\varphi(\ell)$ describing a solenoidal flow-line lying on the $A = A^*$ flux-surface,
\begin{equation} 
\label{flow_line2}
\varphi (\ell, {\bf v}) =  \varphi_0 +{1 \over 2 \pi} \int_0^{\ell} ~  \left[ {\Psi \over \sqrt{\gamma_0^2|\nabla A|^2 + \Psi^2}} \right]_{A = A^*} {d\ell \over r \sin \theta} ,
\end{equation}
\noindent
identifying $\ell = 0$ as the starting-point of the subject flow-line located on the flux surface $A = A^*$ and subtituting $\phi (A) = \gamma_0 A$.   The flow lines are generally a single, ergodic, infinitely-long line filling up the entire flux surface $A = A^*$.   Thus, the flux surface $A = A^*$ is filled densely by a pair of infinitely-long, single, ergodic, field and flow lines that mutually intersect, steadily inducing an electric potential field ${\bf E}$.  On the surface $A = A^*$, three functions, the flux-function $A$, flow-function $\phi = \gamma_0 A$ and electric-potential $W \propto A |A|^{1/2}$ given by equation (\ref{W_1.5A}), take constant values.  

\begin{figure*}
\label{Fig5}
\centerline{\includegraphics[width=60mm]{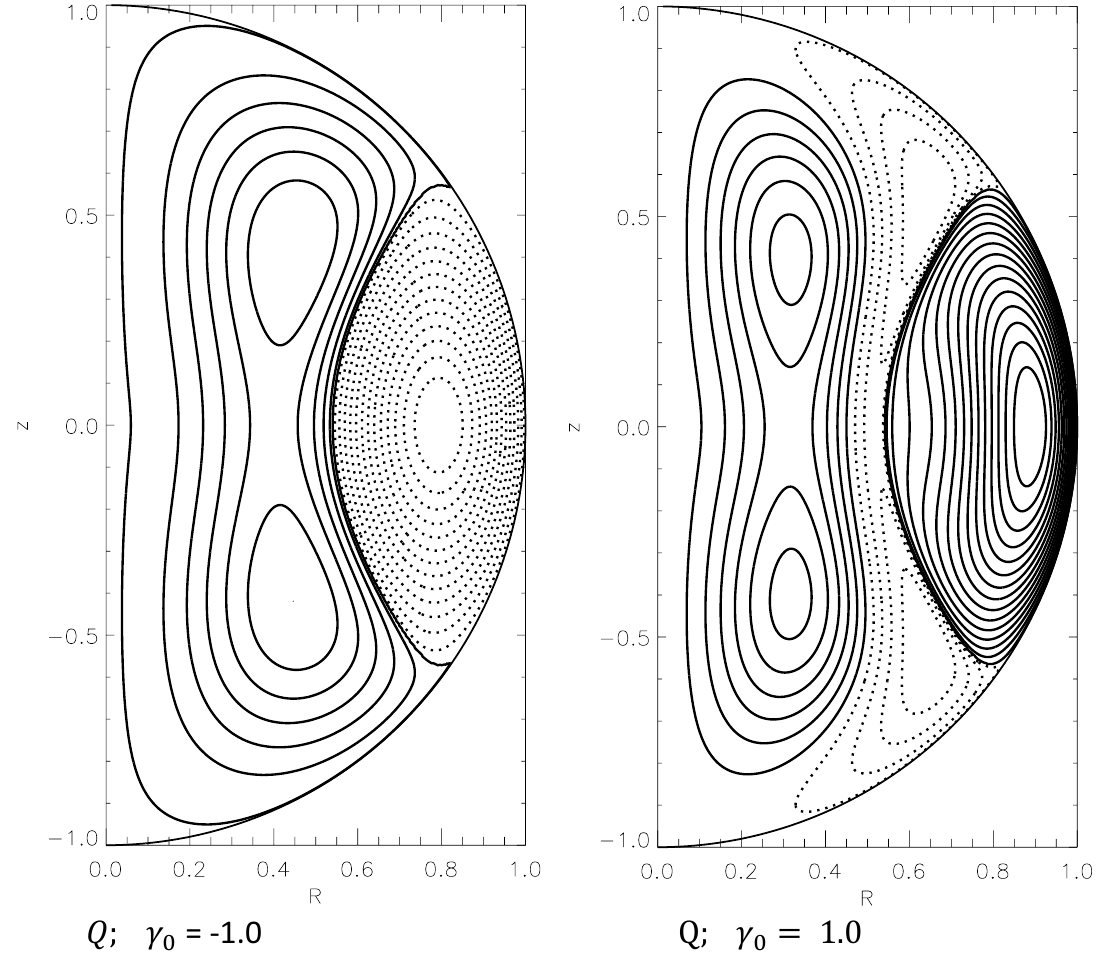}}
\caption{\small{Contour-plots of $Q = r \sin \theta B_{\varphi}$ of the field-aligned Chandrasekhar equipartition steady-flows described by eigenfunction $A^{(m=1)}(r, \theta)$ with $\zeta_0 = \zeta_{4, (m=1)} = 8.1826$, $\gamma_0 = \pm 1.0$, $\lambda_0 = 0$, and ${\bf v} = \pm{{\bf B} \over \sqrt{4 \pi \rho_0}}$.  The mathematical solution $A^{(m=1)}(r, \theta)$ is everywhere physically meaningful, positive and negative respectively in the $V_{(\pm)}$ sub-volumes.}\\ \\ \\}\end{figure*}

To take a different perspective, the steady spatial distribution of ${\bf B}$ is stationary but the fluid is not static.  The field at each point in space at any given time is being bodily carried by a fluid element located at the point in its motion along its steady flow path.   The fluid element is moving with a time-dependent velocity in the Lagrangian sense, that velocity being the steady Eulerian velocity ${\bf v}$ at the instantaneous location of the moving fluid element. The steady velocity ${\bf v}$ crossing the local field ${\bf B}$ induces a dynamo electric-field ${\bf E}$ that regenerates the field in the moving fluid element.  Under the frozen-in condition, the dynamo action takes the form of the flux being conserved as seen by the moving fluid-element.  Relative to the moving fluid-element, its field is changing with time along its flow path, the change dictated by its unchanging frozen-in flux.  In a steady cross-field flow, the field ${\bf B}$ is stationary in space because its global spatial distribution is consistent with the active Lagrangian transport of magnetic flux by every fluid element in the steady flow, the physical essence of the steady dynamo in action.  

In this Subsection, we have concentrated on the $m = 1$ eigenfunction TsF flows, as a sufficient first-principles illustration of similar constructions with the $m > 1$ eigenfunctions.  To complete our analysis of the TsF $m = 1$ cross-field flows, Fig. 5 displays the TsF flow-solutions in the two limit cases of $\gamma_0 = \mp 1$, each an Alfv\`{e}nic Chandrasekhar equipartition flow, with ${\bf v} = \mp {{\bf B} \over \sqrt{4 \pi \rho}}$ in full alignment.  Setting $\gamma_0 = \mp 1$ implies that the azimuthal components of both flow and field are aligned, by equations (\ref{PsiQ10}) and (\ref{PsiQ11}).  The two Chandrasekhar flow-solutions satisfy Tsinganos PDE (\ref{tsing_5}) on two counts, by $\lambda_0 = V_A\left(1 - \gamma_0^2\right) = 0$ and by $A$ satisfying PDE (\ref{tsing_6}).  The prescription $\gamma_0^2 = 1$ alone ensures that the steady-flow equations are satisfied for a Chandrasekhar equipartition flows.  Whereas, $A$ being a solution of PDE (\ref{tsing_6}) is incidental, the two flows picked out by the PDE from the continuum of Chandrasekhar flows with freely prescribable field geometries.  These two flows displayed in Fig. 5 are physically well defined for {\it both} $V_{\mp}$ sub-volumes.   

\subsection{Tsinganos-Prendergast field-aligned flows}

Setting $s_2 = 0$, ${\bf E} = 0$ in Tsinganos PDE (\ref{tsing_6}), 
\begin{equation}
\label{tsing_8}
{\mathcal L} A  +s_1^2 A + p_1 r^2 \sin^2 \theta = 0 ,
\end{equation}
which has the solutions in terms of the variables $\zeta = s_1 r$ and $\zeta_0 = s_1 r_0$, 
\begin{eqnarray}
\label{TsP_A}
A (r, \theta)= a_1 \left[\zeta j_1(\zeta) - {\zeta^2 j_1(\zeta_0) \over \zeta_0} \right] \sin^2 \theta , ~ \zeta < \zeta_0 ,
\end{eqnarray}
\noindent
under the three stress-free boundary conditions (\ref{general_bc2}), describing the Tsinganos-Prendergast \citep{tsinganos1981} field-aligned steady flows in a spherical star of radius $r_0$.  The constant $a_1$ is a freely prescribed amplitude whereas the stress-free the parameter $\zeta_0 = s_1 r_0 = \zeta_{2, m}$ belongs to a discrete spectrum of eigenvalues satisfying the Prendergast\citep{prendergast1956} eigenvalue equation
\begin{equation}
\label{Prendg_eigenvalue}
j_2(\zeta_{2, m}) = 0, ~~~m = 1, 2, 3, ... ,
\end{equation}
\noindent
with $\zeta_{2,m=1} = 5.7635$, $\zeta_{2,m=2} = 9.0950$, citing the first 2 eigenvalues.  For each eigenvalue $\zeta_0 = \zeta_{2, m}$, 
\begin{equation}
\label{Prendf_p_1}
p_1 = s_1^2 a_1 {j_1(\zeta_{2, m}) \over \zeta_{2, m}} , 
\end{equation}
\noindent
which defines the Bernoulli pressure and the fluid pressure given by 
\begin{equation}
\label{P_0_condition2_TsP}
p + \frac{1}{2} v^2 = \rho_0 \left\{ \frac{1}{2} {G M_0\over r_0^3} \left(r_0^2 -  r^2 \right)+ \lambda_0 p_1 A \right\}, ~~~ r \le r_0 , 
\end{equation}
\noindent
obtained from equation (\ref{P_0_condition2}) with $s_2 = 0$. 

\begin{figure*}
\label{Fig6}
\centerline{\includegraphics[width=60mm]{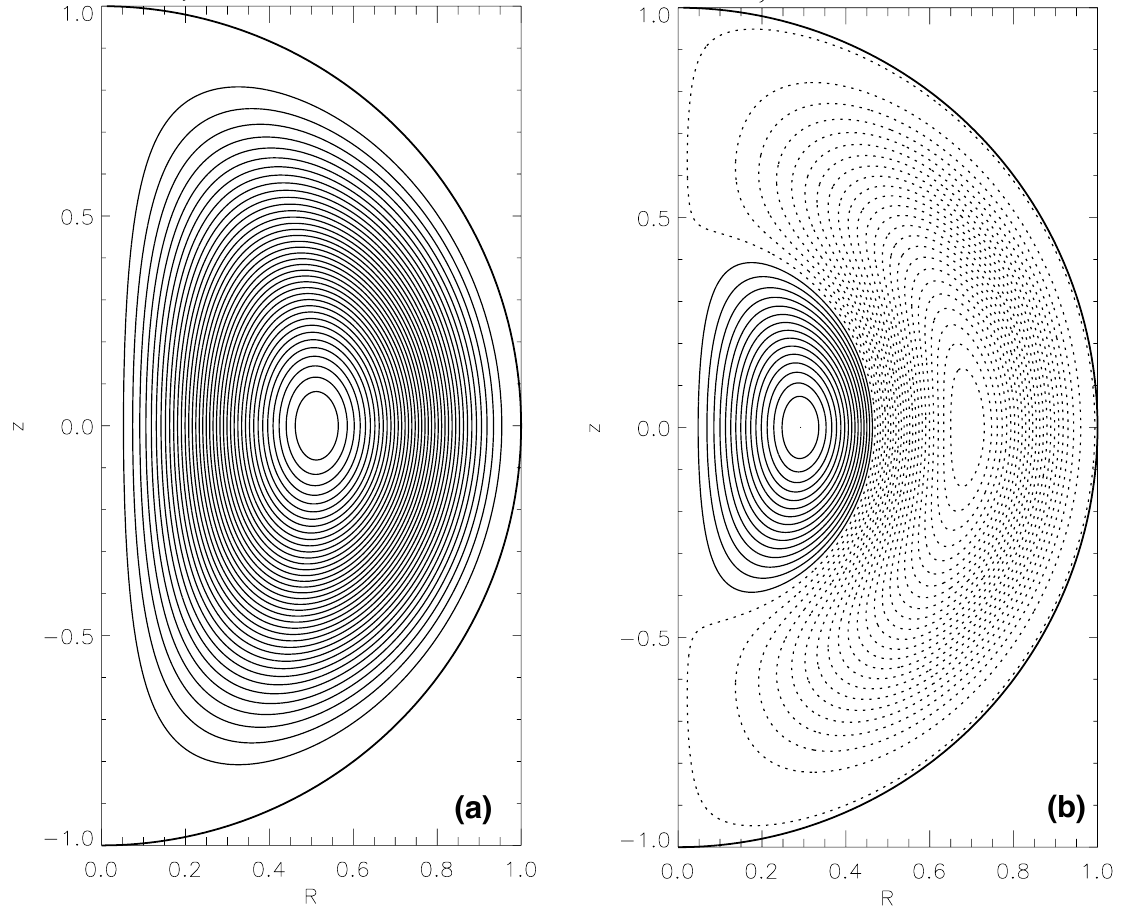}}
\caption{\small{Two axisymmetric, spherically-shaped Tsinganos-Prendergast stellar flows of radius $r_0 = 1.0$ abutting vacuum, represented by the closed contours of their respective flux-function $A(r, \theta)$ in cylindrical coordinates $(R, z) = (r, \theta)$, for the first two eigenvalues (a) $\zeta_{2, m=1} = 5.7635$, on the left, and (b) $\zeta_{2, m=2} = 9.0950$, on the right, in the same format as in Fig. 1c.  In each flow, the velocity ${\bf v}$ and field ${\bf B}$ are uniformly aligned, with ${\bf E} = 0$, both field and flow vanishing at the free-boundary $r = r_0$.  The field-aligned flow lines are generally ergodic on the nested toroidal flux surfaces channelling vortical flows around the symmetry axis.  The two TsP flows respectively comprise a single equator-centered flux-rope, on the left, and two equator-centered flux-ropes, on the right.}\\ \\ \\}\end{figure*}

These solutions describe the continuum of Tsinganos-Prendergast\citep{prendergast1956, tsinganos1981} (TsP) stars with field-aligned flows.  Setting $B_0 = a_1$ fixes the amplitude of the poloidal field ${\bf B}_p$.  Then, setting $v_0 = v_A = B_0 \left( 4 \pi \rho_0 \right)^{-1/2}$ to fix the normalization constants, the poloidal flow ${\bf v}_p$ and field ${\bf B}_p$ are aligned, given by equations (\ref {v_pB_p}) and (\ref {lambda_0}), with $\lambda_0 = v_A^2 \left( 1 - \gamma_0^2\right)$.  The constant $\gamma_0$ then determines the poloidal flow and field, as well as the azimuthal components
\begin{eqnarray}
\label{PsiQ12}
\Psi &=&  \gamma_0 s_1 A, \\ 
\label{PsiQ13}
Q &=& s_1 A ,
\end{eqnarray}
\noindent 
given by equations (\ref{PsiQ8}) and (\ref{PsiQ9}), that is, $v_{\varphi} = \gamma_0 B_{\varphi}$, and it follows ${\bf v} = \gamma_0 {\bf B}$.  The flow is everywhere sub-Alfv\`{e}nic if $\lambda_0 > 0, \gamma_0^2 < 1$ and everywhere super-Alfv\`{e}nic if $\lambda_0 < 0, \gamma_0^2 > 1$.  The case $\gamma_0 = 1$ gives an Alfv\`{e}nic Chandrasekhar equipartition flow.

Figs. 6 displays the solutions $A(r, \theta)$ describing the TsP stars identified by the first two eigenvalues $\zeta_{2,m=1} = 5.7635$, $\zeta_{2,m=2} = 9.0950$.

\section{Relating ideal and near-ideal fluids}

The steady free-boundary flows in an ideal incompressible star is a complete physical system in its simplicity of having the nonlinear Alfv\`{e}n waves as its only magnetic wave phenomenon.  The TDs in ${\bf v}$ and ${\bf B}$ define the ideal fluid physically, how and where TDs form being its fundamental property.  The general first-principles treatment in Section II underlies the axisymmetric steady-flow solutions constructed in Section III.  Here we present several perspectives that clarify the relationship between ideal fluids and the near-ideal fluids of astrophysics, thinking physically when nonlinear intractability stands in the way.  Central to the narrative is the question how steady flows may physically arise, those in a rigorously ideal fluid quite distinct from those in a near-ideal fluid with its hallmark propensity for viscous and resistive dissipation via near-TDs\citep{woltjer1958, parker1972, parker1991,yu1973, berger1984, taylor1974, bergerField1984, parker1994, boo2010, low2011, boo2014, lf2014, gp2016, low2023}.

\subsection{Parker spontaneous current sheets}

We are not aware of any published work treating the finite-time singularities of 3D PDE (\ref{momentum3}) describing incompressible Alfv\`{e}n waves, but the nature of these singularities can be intuitively related to the magnetic frozen-in condition.  Mass is conserved between any pair of frozen-in flux surfaces, the flux volumes  ready to slip frictionlessly and discontinuously along their boundaries where magnetic and flow TDs form.  In 3D dynamics, two spatially separate flux-volumes can make contact over a newly-created common flux-surface, by pushing their ways through an intervening layer of sheared field.  The resulting complex of deformed flux surfaces would contain TDs wherever two magnetically-unconnected layers of field have mutually slipped.  The slipping is ideal and reversible, with no breaking and rejoining of field lines, notwithstanding the holes having been punched through layers of flux surfaces.  Each hole is created in a flux surface by the discontinuous parting of field lines in the surface. 

Forming and dissolving flux-surface holes are neatly described by an optical analog\citep{parker1991}, the field lines in a flux surface behaving as optical rays governed by a refractive index proportional to $|{\bf B}|^{1/2}$.  Field lines are refracted concave to a local maximum in $|{\bf B}|^{1/2}$ and if the maximum is sufficiently localized in the flux surface, field lines are excluded completely from a finite area around the maximum.  The exclusion area is then a hole in the flux surface that can close up when the maximum dissolves away as the flow evolves. The ideal PDEs (\ref{momentum})-(\ref{Newton}) describing an initially everywhere continuous flow loses analyticity upon the formation of the first TD or flux-surface hole, a property of the governing hyperbolic PDEs.  

The 3D fluid displacements creating TDs are driven by a peculiar geometric incompatibility among the  forces in momentum PDE (\ref{momentum3}), hereafter referred to as the three forces.  The centrifugal and magnetic-tension forces at each point in space are confined to their respective osculating planes of the flow-line and field-line passing through the point, each plane defined by the tangent and normal of the line at the point.  Whereas, the third force is directed along the normal to the level surface of the total pressure $P_T$ where the point is located.  The three forces cannot be freely determined at each point in space, being globally subject to the invariance of topology ${\mathcal T}$.  In a time-dependent state, the non-equilibrium distribution of the three forces at any time is correlated nonlinearly across all space by propagating Alfv\`{e}n waves.  The formation of TDs is an essential local response of the fluid to the globally correlated, anisotropic forcing, resulting in frictionless, discontinuous slipping of magnetic volumes of fluid along flux surfaces wherever such a slipping is energetically favored\citep{parker1994, low2023} .  From this perspective, demanding analyticity for all time is seen as physically extraneous to the inviscid, perfectly conducting fluid. 

\subsection{The phase-spaces ${\mathcal B}_{\mathcal T}$ and ${\mathcal B}^*_{\mathcal T}$}

An ideal incompressible hydromagnetic star has a fixed entire history determined by its permanent physical properties. Incompressibility constrains the stellar fluid to move via volume-preserving displacements with a given uniform density $\rho_0$.  The fixed stellar volume $V_0$ defines the unchanging total mass $M_0 = V_0 \rho_0$ as its free boundary $\partial V$ deforms.  The stellar magnetic field ${\bf B}$ has a fixed topology ${\mathcal T}$ under the frozen-in condition, {\it permanently} partitioning the fluid into continua of specific sub-volumes identifiable at any time in the fluid.

To keep essential concepts simple, we limit attention to the field being wholly contained in the fluid abutting vacuum with a conserved total energy ${\mathcal E} < 0$.  The energy ${\mathcal E}$, as the sum of magnetic, kinetic and gravitational energies, being negative is a statement of gravitational confinement.  Although the incompressible fluid cannot expand to infinity, the confinement may be unstable to breaking up into separate stellar bodies orbiting around each other in mutual gravitational confinement conserving ${\mathcal E} < 0$.  It is also possible for a sub-body to be ejected out to infinity, leaving the remaining orbiting sub-bodies at a total energy less than ${\mathcal E}$.  Hydromagnetic instability may extend an interior field out of the star into a potential field, which reduces the total energy ${\mathcal E}$ by the escape of Maxwell’s vacuum electromagnetic waves out to infinity\citep{low1982}.  We keep the preceding ramifications in mind as we construct the permanent physical identity of a single stellar fluid abutting vacuum, conserving the total energy ${\mathcal E}$ and preserving magnetic topology ${\mathcal T}$.

The single stellar fluid is defined for all time by its incidental state at any time $t = t_0$, in terms of its initial shape $\partial V_{t = t_0}$ together with initial flow/field $\left[{\bf v} \left({\bf x}, t_0 \right), {\bf B} \left({\bf x}, t_0 \right)\right]$.  This initial state defines the conserved ${\mathcal E}$ and invariant topology ${\mathcal T}$.   Since the governing hydromagnetic equations are invariant under time-reversal symmetry $t \rightarrow -t, {\bf v} \rightarrow - {\bf v}$, the given initial conditions define the evolution of the stellar fluid for all time.  

Let us assume that the flow and field are continuous in space for all time, the assumption to be relaxed later in the analysis.  We define the phase-space ${\mathcal B}_{\mathcal T}$ to contain all the continuous fields ${\bf B}$ of topology ${\mathcal T}$, arbitrarily and continuously deformed from ${\bf B} \left({\bf x}, t_0 \right)$, just a mathematical construction separate from ideal flows governed by the hydromagnetic equations.  The entire $(-\infty < t < \infty)$ history of the hydromagnetic flow of the given fluid identified by the $t = t_0$ initial state is then represented by the flow’s magnetic field ${\bf B} \left({\bf x}, t \right)$ tracing an evolutionary path in the phase space ${\mathcal B}_{\mathcal T}$, hereafter called a world-line.  

The world-line constructed with ${\bf B} \left({\bf x}, t_0 \right) \in {\mathcal B}_{\mathcal T}$ is determined by the initial velocity ${\bf v} \left({\bf x}, t_0 \right)$ in accordance to the ideal hydromagnetic equations.  Different initial velocities prescribed at time $t = t_0$ for the same ${\bf B} \left({\bf x}, t_0 \right)$ define different world-lines originating from ${\bf B} \left({\bf x}, t_0 \right) \in {\mathcal B}_{\mathcal T}$.  A world-line so identified by the $t = t_0$ initial state is also identified equivalently by the state $\left[\partial V, {\bf v} \left({\bf x}, t \right), {\bf B} \left({\bf x}, t \right)\right]$ at any time $t \ne t_0$ along the world-line.  The world-lines catalog the entire continuum of time-dependent hydromagnetic evolutions of a fluid identified by its given invariant field-topology ${\mathcal T}$.  

The spontaneous formation of the Parker current sheets shows that generally, the world-lines in the 3D dynamics cannot remain analytical and must leave the phase-space ${\bf B} \left({\bf x}, t \right) \in {\mathcal B}_{\mathcal T}$.   TD formation is ideally reversible.  Therefore, phase-space ${\mathcal B}_{\mathcal T}$ must be contiguous with the phase-space ${\mathcal B}^*_{\mathcal T}$ containing fields with TDs of the invariant topology ${\mathcal T}$, arbitrarily and mathematically deformed from each continuous field in ${\mathcal B}_{\mathcal T}$.  Thus, the union of the two phase-spaces ${\mathcal B}_{\mathcal T} \cup {\mathcal B}^*_{\mathcal T}$ contains all admissible world-lines of a given fluid of a fixed field topology ${\mathcal T}$ including evolutions containing TDs.  A world-line may continue from ${\mathcal B}_{\mathcal T}$ into ${\mathcal B}^*_{\mathcal T}$ with the first TD formation, or, continue from ${\mathcal B}^*_{\mathcal T}$ into ${\mathcal B}_{\mathcal T}$ upon the dissolving of the last TD to render ${\bf v}$ and ${\bf B}$ continuous in space.  

The steady flows of a fixed ${\mathcal T}$ are represented by isolated stationary points in ${\mathcal B}_{\mathcal T} \cup {\mathcal B}^*_{\mathcal T}$ with no world-lines to speak of.  The Parker\citep{parker1994} theory stating that TD-bearing flows are the rule rather than the exception, may be reworded to claim that 3D steady flows depending on its topology ${\mathcal T}$ are mainly located in phase-space ${\mathcal B}^*_{\mathcal T}$ rather than ${\mathcal B}_{\mathcal T}$.  The concept of “mainly” applied to the continua of steady-flow stationary points can be rendered precise using the probabilistic description of measure theory\citep{schwartz1966}.  The steady-flow stationary points located in ${\mathcal B}^*_{\mathcal T}$ are “immensely greater in number” than the steady-flow stationary points located in ${\mathcal B}_{\mathcal T}$ in the following sense.  Randomly selecting a field topology ${\mathcal T}$, the probability of finding an everywhere continuous steady-flow in ${\mathcal B}_{\mathcal T} \cup {\mathcal B}^*_{\mathcal T}$ is nil.  That is, the everywhere continuous steady-flows in ${\mathcal B}_{\mathcal T}$ are sparsely distributed in the combined phase-space ${\mathcal B}_{\mathcal T} \cup {\mathcal B}^*_{\mathcal T}$.  

A instructive analogy is that randomly picking a number in the open unit interval between 0 and 1 that happens to be a rational fraction of two integers, is nil.  The infinitely many, {\it countable} rational numbers are sparsely distributed amidst the infinitely many, {\it uncountable} irrational numbers in the interval. 

There is a wealth of formidable topological structures of the phase-space ${\mathcal B}_{\mathcal T} \cup {\mathcal B}^*_{\mathcal T}$ which lies outside the scope of our study.  The world-lines in the ${\mathcal B}_{\mathcal T} \cup {\mathcal B}^*_{\mathcal T}$ are likely ergodic by a suitable metric definition of closeness between two points in the phase-space, embedding sparsely distributed closed world lines of finite path lengths describing periodic evolutions.  More common are the quasi-periodic ergodic evolutions, each described by a never-ending world line confined in a bounded, multiply-connected, multi-dimensional subspace of ${\mathcal B}_{\mathcal T} \cup {\mathcal B}^*_{\mathcal T}$, such as described by the Kolmogorov-Arnold-Moser Theorem\citep{Arnold1997} in Hamiltonian dynamical systems.   

\subsection{Linear and nonlinear Kelvin-Helmholtz instabilities}

Vorticity ${\bf w} = \nabla \times {\bf v}$ is not conserved in the presence of the field ${\bf B}$ because the angular momentum in a parcel of fluid gains and loses by the Lorentz force.  It is the frozen-in flux surfaces that define the permanent sub-volumes, each moving as a whole and capable of tangentially slipping discontinuously along another.  The limit case of a fluid without a field is interesting in its own right, for which momentum equation (\ref{momentum4}) reduces to  
\begin{equation}
\label{Kelvin}
{\partial {\bf w} \over \partial t} + \nabla \times \left( {\bf w} \times {\bf v} \right) = 0 ,  
\end{equation} 
\noindent   
describing the vorticity ${\bf w} = \nabla \times {\bf v}$ to be frozen in the fluid.  The frozen-in vorticity was discovered by Kelvin\citep{saffman1992} before the frozen-in field was recognized by Alfv\`{e}n\citep{AlfvenFalthammar1963} as an analogous property.  In the absence of the magnetic field, the inviscid fluid moves with an invariant vorticity topology.  The vortex surfaces permanently partition the fluid into sub-volumes that can tangentially slip frictionlessly and discontinuously.  Thus, the Parker theory may be reworded to state that everywhere-continuous inviscid flows are the exception rather than the rule, a property underpinning the Kelvin-Helmholtz instabilities.

Velocity shears in an ideal fluid are linearly unstable\citep{saffman1992}.  Taking Kelvin-Helmholtz instabilities beyond linear analysis into weakly nonlinear regimes had revealed a loss of analyticity\citep{moore1979, calflischorellana1986}.  Not pointed out in these pioneer analyses, the encountered singularities are suggestive of an irrepressible, frictionless and discontinuous slipping among vorticity volumes via energetically favored fluid displacements.  Arnold\cite{arnold1965, moffatt1985} pointed out that vortex-sheet singularities arise naturally in the ideal non-magnetic fluid.  The basic idea thus emerges that the probability is nil for finding a 3D spatially-continuous steady flow possessing a randomly-picked, invariant vorticity topology.       

\subsection{Astrophysical near-ideal incompressible fluids}

Consider a dissipative fluid with weak viscosity and electrical resistivity represented by constant coefficients $\nu$ and $\eta$, respectively, in its parabolic governing PDEs
\begin{eqnarray}
\label{momentum_viscous}
\rho_0 {\partial {\bf v} \over \partial t} + \rho_0 ( {\bf v} \cdot \nabla ) {\bf v}  &=& \frac{1}{4 \pi} \left( \nabla \times {\bf B} \right) \times {\bf B} - \nabla p \nonumber \\
&& ~~~~~~ - \rho_0 \nabla U + \nu \nabla^2 {\bf v} ,\\
\label{induction_resistive}
{\partial {\bf B} \over \partial t} &=& \nabla \times \left( {\bf v} \times {\bf B} \right) + \eta \nabla^2 {\bf B} .
\end{eqnarray}
\noindent
Dissipation of kinetic and magnetic energies takes place as diffusion of ${\bf v}$ and ${\bf B}$ controlled by coefficients $\nu$ and $\eta$.  Pressure remains an {\it in situ} reaction force governed by elliptic Poisson PDE ({\ref{p_Poisson}) whereas the incompressible Alfv\`{e}n waves are damped by resistivity and viscosity.  Diffusion takes place at all speeds, so that gradients in ${\bf v}$ and ${\bf B}$ diffuse faster than possible for discontinuities to form.  

Astrophysical viscosity and electrical resistivity are not zero, but are significant only over length-scales orders of magnitudes smaller than astronomical length-scales.  The dimensionless Reynolds number ${\mathcal R} = {v_0 L_0 \rho_0\over \nu}$ and Lundquist number ${\mathcal L} = {v_0 L_0 \over \eta}$ respectively compare a macroscopic velocity $v_0$ against the diffusion speeds of viscosity and resistivity over a macroscopic length-scale $L_0$.  The common astrophysical environment with ${\mathcal R} >> 1$, ${\mathcal L} >> 1$ indicates  that, provided a flow does not incur extreme gradients on its own, a near-ideal fluid behaves essentially as an ideal fluid.  The Parker theory posits that unlimitedly small scales almost always develop nonlinearly in a 3D ideal fluid to the point of TDs forming and persisting in its flow, if viscous/resistive dissipation is absolutely suppressed.  Radically different, a 3D near-ideal fluid develops irrepressible near-TDs on macroscopic scales with the near-TDs thinning to the point of being dissipated by the otherwise weak viscosity/resistivity.  The total energy ${\mathcal E}$ decreases stochastically with the incidental dissipation of near-TDs as the field topology ${\mathcal T}$ changes.  

Although magnetic reconnections\citep{parker1979, parker1994, kulsrud2005, gp2016} take place on the small scales, each topological change from ${\mathcal T}$ to ${\mathcal T}’$ is global.  Local field-line breaking and reconnecting propagate as Alfv\`{e}n waves in the changed field topology ${\mathcal T}’$ as the free boundary $\partial V$ adjusts.  The near-ideal evolutionary world-line ends in ${\mathcal B}_{\mathcal T} \cup {\mathcal B}^*_{\mathcal T}$ and continues in ${\mathcal B}_{\mathcal T’} \cup {\mathcal B}^*_{\mathcal T’}$. Each reconnection episode simplifies the field topology and removes kinetic and magnetic energies.  The weaker viscosity/resistivity happen to be, the more effective is a 3D near-ideal fluid capable of macroscopically creating near-TDs for dissipation.   The episodic reconnections repeat extensively in the fluid, simplifying field topology and decreasing total energy ${\mathcal E} < 0$ in an ongoing, irrepressible, stochastic relaxation.  

The near-ideal evolution is thus a world-line traced in the combined phase-space summed over all admissible topology ${\mathcal T}$
\begin{equation}
{\mathcal B}_{all} = \bigcup_{\mathcal T} ~ \left[ {\mathcal B}_{\mathcal T} \cup {\mathcal B}^*_{\mathcal T} \right] .
\end{equation}
\noindent
The relaxation process in ${\mathcal B}_{all}$ naturally seeks a terminal minimum-${\mathcal E}$ steady state, driven by the Parker current-sheets forming and dissipating in a 3D field, never ending because the probability is nil for arriving in a rigorously TD-free steady state.  However, the irrepressible current sheets must form with diminishing current intensities as free energy runs out by the simplification of field topology ${\mathcal T}$.  The fluid thus naturally relaxes into an energetically preferred macroscopic steady-flow, with TDs still forming and dissipating but at progressively insignificant energy dissipations.  The axisymmetric flows in Section III if linearly stable may thus be viewed as the products of such a stochastic relaxation of a 3D flow.

The gravitating, incompressible, {\it non-magnetic} 3D stars isolated in vacuum presents a corollary.  Taking viscosity to be rigorously zero, the fluid evolves in the sub-phase-space defined by its absolutely invariant vorticity-topology, generally embedding irrepressible, infinitesimally-thin vortex sheets.  Whereas, a near-inviscid fluid dissipates such vortex sheets, changing its vorticity-topology and evolving openly in the combined phase-space of all vorticity-topologies.  This turbulent behavior is intriguing in the evolution of a collection of interacting near-ideal isolated vorticities\citep{McWilliamsPlierl1979}. 
 
\subsection{Hydromagnetic dynamo action}

The turbulent relaxation process of a near-ideal fluid is distinct from an astrophysical field actively generated by turbulent dynamo actions such as occurring in the Sun as a magnetic variable star\citep{parker1979,evs2019}.  In broad terms, the Sun rotates, with a core possibly in uniform rotation, that is enveloped by a differentially-rotating, turbulently-convective outer shell of an estimated thickness of about a fifth of a solar radius\citep{CaligariMoreno-InsertisSchussler1995, yfan2009,HesterZhangDikpati2025}.  A solar dynamo maintains the Sun’s dipolar, global magnetic field that reverses polarity approximately every eleven years\citep{parker1979, ChoudhuriSchusslerDikpati1995,DikpatiCharbonneau1999, yfanfang2014, McIntoshLeamon2024, McIntoshLeamonEgeland2019}.  This dynamo is driven by the hydromagnetic flows delivering the imperative outpouring of thermonuclear energy from the core into the convective envelope, to escape passively as white light through the optically-thin solar atmosphere, the corona, at a highly steady rate of about $4 \times 10^{33} ~ erg~ s^{-1}$.   Maintaining the luminosity throughput dominates the solar interior and is probably the dynamical reason for the steadiness of the Sun’s highly spherical shape.  The solar dynamo field is energetically driven, a 3D process, the large-scale field sheared by differential rotation and subject to convection-scale twisting and folding of magnetic fluxes and vorticity.  Magnetic topological change via resistive field-diffusion is essential in the dynamo folding of magnetic flux\citep{parker1979}.  The solar dynamo is, of course, beyond the scope of our basic study.  However, there is a fundamental dynamo effect not previously recognized, about which our study suggests an interesting insight.

The dynamo interaction between magnetic twist and vorticity may self-arrest or be quenched by the creation of meta-stable field-aligned and cross-field steady flows as products of a global dynamo, of the kind represented by the axisymmetric steady flows in Section III.  The inexorable near-ideal drive to locate a terminal steady-flow as a dynamical attractor, is brought about by free energy running out as irrepressible TDs form with diminishing free energy.  That a Tsinganos-Ferraro steady cross-field flow can exist is remarkable.  Its mutually intersecting pairs of infinitely-long, ergodic, solenoidal flow-lines and field-lines, fill up entire flux surfaces.  Such a topologically complex steady hydromagnetic flow seems possible only as a product of turbulently-dissipative self-organization\citep{taylor1974, berger1984}.  

That dynamo actions may be locally quenched in this manner is intriguing.  Dynamo actions cease locally when the flow-lines and field-lines are forced into alignment by the turbulent resistive diffusion of the field and dissipation of vorticity, or, when a steady Tsinganos-Ferraro cross-field flow emerges because it is energetically favored.  Storage of steady vorticity and magnetic twist in localized pockets may break away from the global dynamo, to coalesce into larger flows or return to be a part of global dynamo actions.  A storage of vorticity and magnetic twist may last long enough to rise and break through the solar surface as the general origin of sunspots\citep{Hathaway2015, Lites1995}.  

Our study was motivated by the intriguing small-scale coronal eruptions\citep{mcintosh2017} persisting coherently along solar longitudes recently observed by STEREO and SDO spacecraft in the period 2011-2013. These eruptions possibly also have their origin in fluxes emerging into the corona, a million-degree hot, near-ideal, spatially-enormous, tenuous external atmosphere.  The eruptions are local  ${\mathcal R} >> 1$, ${\mathcal L} >> 1$ reconnection events as the the emerging vorticity and twisted-fields adjust in the electrically-highly conducting corona.  The eruptions may also be a signature of ongoing spontaneous reconnections that re-configure a large-scale, highly-structured coronal flux-system\citep{gibsonfanmandrini2004, Raouafi2009, ZhangFlyerLow2006, Berger2011, Chen2025} that eventually loses self-confinement to be expelled as one of the daily Coronal Mass Ejections (CMEs)\citep{gibsonLow1998, fangibson2004, manchest2004, zhanglow2005, fangibson2007} traveling out in the interplanetary solar wind\citep{parker1963}.  

These coronal phenomena have the following fundamental implication.  Coronal flux systems, as self-organized products of ubiquitous flux emergence, are the accumulations of magnetic fluxes lost by the solar dynamo that eventually are taken out by CMEs into interplanetary space.  Which is to say, the solar dynamo is an open system, bodily delivering generated flux into the open solar atmosphere as the basic mechanism of the eleven-year polarity-reversals of the coronal global field \citep{low1997, Gopal2003}.  

\subsection{Flux emergence and ejection from an incompressible star}

Coronal phenomena\citep{low2019} are an essential component of the hydromagnetic Sun.   The loss of magnetic flux and twists from its dynamo is, in fact, also a property of the incompressible star we have studied.  Limiting our treatment to wholly contained stellar fields keeps the physical issues simple and their mathematics manageably tractable.  Whereas, wholly contained fields are generally unstable at the free boundary $\partial V$ abutting vacuum.  The fields are liable to break through $\partial V$ to extend into the vacuum exterior, the emerged field instantanly becoming potential in the non-relativistic hydromagnetic description.  The leading-order description of the vacuum is relativistic, with electric and magnetic fields governed by Maxwell’s equations.  As the incompressible star evolves with a shape-changing $\partial V$ and an instantaneously-changing emerged potential field, two consequences are fundamental.  

As previously pointed out, total energy ${\mathcal E}$ is then no longer conserved and must be decreasing in time, the lost energy carried away by the Maxwell waves neglected in the hydromagnetic description.  The incompressible fluid may remain whole as the star, but magnetic twist must travel out along field lines crossing $\partial V$ to also be carried away by Maxwell waves.  The field topology ${\mathcal T}$ is no longer preserved.  Thus, at any one time during flux emergence, the external potential field behaves as a sink of energy, magnetic fluxes and magnetic helicity as a measure of magnetic twists.  The whole physical system including the exterior vacuum can be described by coupling Maxwell equations to the non-relativistic hydromagnetic equations governing the stellar interior, such as seen in a recent calculation\citep{low1982}.   The Maxwell waves transporting energy and magnetic flux and helicity to infinity correspond to the hydromagnetic outward transport of similar losses to the solar corona via the CMEs\citep{Low1994, Rust1994}.  

The theoretical motivation of our study bears reminding, to discover hydromagnetic properties in the simplicity of the incompressible fluid possessing a single magnetic-wave process.  These properties are interesting in their own right as a part of continuum mechanics while, as our study shows, they offer generalizable conceptual ideas for interpreting the observed Sun.   The study calls for a theoretical effort to develop 3D time-dependent Direct Numerical Simulation computation codes capable of treating the elliptic, hyperbolic and parabolic hydromagnetic PDEs, building on ongoing numerical studies\citep{DStLowthesis,desterckpodts1999, KerrBrandenburg1999, GrafkeHomannDreher2008, ramitlowsmolar2010, yfanfang2014, stone1999}.  The success of such a program depends on insightful forethoughts on the relevant physical processes and their nonlinear mathematical properties, to guide and check on the reliability of the codes developed.  

The following  theoretical 3D model holds promise of interesting physics.  As an analog model of the Sun, consider an evolving incompressible star powered by the steady, outwardly-conducting thermal flux from an artificial point-source at the stellar center, that is convectively transported to be radiated away at a variable $\partial V$ into vacuum, across which hydromagnetic structures from a time-dependent interior dynamo continually emerge into outward propagating Maxwell waves.   
 
\section{Concluding remarks} 

The study focused on a broad-brush understanding of hydromagnetic properties, leaving the interested readers to follow up on interesting physical issues encountered in the construction.  The simplicity of the Alfv\`{e}n waves being the only magnetic wave-phenomenon in the incompressible fluid suggests that a mathematical proof of the Parker spontaneous current sheets may be based on the hyperbolic nature of the incompressible PDE (\ref{momentum3}).    

To cite a class of problems one might pursue with physical forethoughts, the linearly-stable, equipartition, field-aligned steady flows of Chandrasekhar admit fields of all topologies, with the flow speed everywhere equal to the Alfv\`{e}n speed.  Thus, each parcel of fluid sees no propagating Alfv\`{e}n waves in its rest frame\citep{parker1979}.  Whereas, in a Tsinganos-Prendergast field-aligned flow, a parcel of fluid in its rest frame sees Alfv\`{e}n waves propagating in all directions or in swept-back directions, depending on whether the flow speed is sub-Alfv\`{e}nic or super-Alfv\`{e}nic.  How do these wave properties affect the stability or instability of the field-aligned flows?  Can an unstable super-Alfv\`{e}nic flow in a near-ideal fluid be driven and then nonlinearly arrested or quenched by TD dissipation into a 3D Chandrasekhar equipartition flow in its linear stability?  All but a few of the field and flow lines of the Tsinganos-Ferraro steady flows have no ends, infinitely-long lines, each line densely covering a whole flux surface.  These steady flows may emerge from  TDs forming and dissipating by magnetic reconnections, densely distributed throughout a fluid\citep{hk1985, low2007}.  

\section{Acknowledgments}  BC Low as Visiting Scientist thanks Holly Gilbert, Director, and Mike Wiltberger, Interim Director, of High Altitude Observatory for support.  This material is based upon work supported by the NSF National Center for Atmospheric Research, which is a major facility sponsored by the US National Science Foundation under Cooperative Agreement No. 1852977.  SW McIntosh appreciates the support of Lynker through its Internal Research and Development Program and sponsorship of his contribution to this work.  This manuscript was prepared using the AIPsubstyles for REVTeX 4.2 Copyright \copyright 2014 by American Institute of Physics.

% If in two-column mode, this environment will change to single-column format so that long equations can be displayed. 
%\begin{widetext}
%$$\mbox{put long equation here}$$
%\end{widetext}

% Figures should be put into the text as floats. 
% Use the graphics or graphicx packages (distributed with LaTeX2e).
% Put the label that you will use with \ref{} command in the braces of the \label{} command.
%
% \begin{figure}
% \end{figure}

% \begin{table}
% \end{tabular}
% \end{table}

% If you have acknowledgments, this puts in the proper section head

%\bibliography{your-bib-file}

\begin{thebibliography}{}

\bibitem[]{low1982} Low, B. C., Self-similar magnetohydrodynamics. II - The expansion of a stellar envelope into a surrounding vacuum, {\it ApJ} 261, 351, 1982.
\bibitem[]{AlfvenFalthammar1963}Alfv\`{e}n, H., \& C.-G. Falthammar, Cosmical Electrodynamics, {\it Oxford U. Press} 1963.
\bibitem[]{LL1960} Landau, L. D., \& E. M. Lifshitz, {\it Electrodynamics of Continuous Media} (Pergamon Press), 1960.
\bibitem[]{parker1979} Parker, E. N., {\it Cosmical Magnetic Fields} (Oxford Univ. Press), 1979.
\bibitem[]{kulsrud2005} Kulsrud, R. M., {\it Plasma Physics for Astrophysics} (Princeton University Press), 2005.
\bibitem[]{parker1972} Parker, E. N., Topological dissipation and the small-scale fields in turbulent gases, {\it ApJ} 174, 499, 1972.
\bibitem[]{parker1994} Parker, E. N., {\it Spontaneous Current Sheets in  Magnetic Fields} (Oxford Univ. Press), 1994.
\bibitem[]{jlp2010} Janse, \AA. M., B. C. Low \& E. N. Parker, Topological complexity and tangential discontinuity in magnetic fields, {\it Phys. Plasmas} 17, 092901,  2010.
\bibitem[]{syrovatskii1981} Syrovatskii, S. I., Pinch sheets and reconnection in astrophysics, {\it Ann. Rev. Astron. Astrophys.}, 19, 163, 1981.
\bibitem[]{hk1985} Hahm, T. S., \& R. M. Kulsrud, Forced magnetic reconnection, {\it Phys. Fluids} 28, 2412, 1985.
\bibitem[]{ZweibelBoozer1985} Zweibel, E. G., \& A. H. Boozer, Evolution of twisted magnetic fields, {\it ApJ} 295, 642, 1985.
\bibitem[]{low2007} Low, B. C., On the possibility of electric-current sheets in dense formation, {\it Phys. Plasmas} 14, 122904 2007.
\bibitem[]{low2015} Low, B. C., Field topologies in ideal and near-ideal magnetohydrodynamics and vortex dynamics, {\it Sci. China Phys. Mech. Astron.} 58, id5626, 2015
\bibitem[]{low2019} Low, B. C., Chapter 6.  Coronal magnetism as a universal phenomenon, {\it The Sun as a Guide to Stellar Physics}, edt. by O. Engvold, J.-C. Vial \& A. Skumanich, (Elsevier), 2019.
\bibitem[]{low2023} Low, B. C., Topological nature of the Parker magnetostatic theorem, {\it Phys. Plasmas}
30, id.012903, 2023.
\bibitem[]{tsinganos1981} Tsinganos, K. C., Magnetohydrodynamic equilibrium. I - Exact solutions of the equations, {\it ApJ} 245, 764, 1981.
\bibitem[]{tsinganos1984} Tsinganos, K. C., R. Rosner \& J. Distler, On the topological stability of magnetostatic equilibria, {\it ApJ} 278, 409, 1984.
\bibitem[]{zaslavskii1988}Zaslavskii, G. M., R. Z. Sagdeev, \& A. A. Chernikov, Stochastic nature of streamlines in steady-state flows, {\it Sov. Phys. JETP} 67, 270,1988. 
\bibitem[]{mcintosh2017} McIntosh, S. W., W. J. Cramer, M. M. Pichardo, \& R. J. Leamon, The detection of Rossby-like waves on the Sun, {\it Nat. Astron.} id0086,  2017.
\bibitem[]{chandra1956} Chandrasekhar, S., On the stability of the simplest solution of the equations of hydromagnetics, {\it Proc. Nat. Acad. Sci.} 42, 273, 1956.
\bibitem[]{chandra1961} Chandrasekhar, S., {\it Hydrodynamic and hydromagnetic stability} (Oxford Univ. Press), 1961.
\bibitem[]{prendergast1956} Prendergast, K. H.,The equilibrium of a self-gravitating incompressible fluid sphere with a magnetic field. I.  {\it ApJ} 123, 498,1956.
\bibitem[]{ferraro1937} Ferraro, V. C. A., The non-uniform rotation of the Sun and its magnetic field {\it MNRAS} 97, 458, 1937.
\bibitem[]{ch1960} Courant, R., \& D. Hilbert, {\it Methods of Mathematical Physics}, Vol. 2, p. 635 (Interscience),1960.
\bibitem[]{schwartz1966} Schwartz, L., {\it Mathematics for the Physical Sciences}, (Hermann, Editeurs des Sciences et des Arts), 1966.
\bibitem[]{cf1976} Courant, R., \& K. O. Friedrichs, {\it Supersonic Flow and Shock Waves} (Springer-Verlag), 1976.
\bibitem[]{Rosner1989} Rosner. R., B. C. Low, K. Tsinganos, M. A. Berger, On the relationship between the topology of magnetic-field lines and flux surfaces, {\it Geophys. Astrophys.  Fluid Dyn.} 48, 251, 1989.
\bibitem[]{dombre1986} Dombre, T., U. Frisch, J. M Greene, et. al., Chaotic streamlines in the ABC flows, {\it J. Fluid Mech.} 167, 353,1986.
\bibitem[]{woltjer1958}  Woltjer, L., A theorem on force-free magnetic fields, {\it Proc. Nat. Acad. Sci.}, 44, 489,1958.
\bibitem[]{yu1973} Yu, G., Hydrostatic equilibrium of hydromagnetic fields, {\it ApJ} 181, 1003, 1973.
\bibitem[]{boo2010} Boozer, A., Mathematics and Maxwell’s equations, {\it Phys. Plasmas} 52, id124002 2005,2010
\bibitem[]{boo2014} Boozer, A., Formation of current sheets in magnetic reconnection, {\it Phys. Plasmas} 21, id072907, 2014.
\bibitem[]{taylor1974} Taylor, J. B., Relaxation of toroidal plasma and generation of reverse magnetic fields, {\it Phys. Rev. Letts}, 33, 1139, 1974.
\bibitem[]{berger1984} Berger, M. A., Rigorous new limits on magnetic helicity dissipation in the corona, {\it Geophys. Astrophys. Fluid Dyn.} 30, 79, 1984.
\bibitem[]{bergerField1984} Berger, M. A., \& G. B. Field, The topological properties of magnetic helicity, {\it J. Fluid Mech.} 147, 133, 1984.
\bibitem[]{low2011} Low, B. C., Absolute magnetic helicity and the cylindrical magnetic field, {\it Phys. Plasmas} 18, 052901, 2011.
\bibitem[]{lf2014} Low, B. C., \& F. Fang, Cylindrical Taylor states conserving total absolute magnetic helicity, {\it Phys. Plasmas} 21, 092116, 2014.
\bibitem[]{gp2016} Gonzalez, W., \& E. Parker (edts.), {\it Magnetic reconnection: Concepts and applications, astrophysics and space science} (Springer), 2016.
\bibitem[]{parker1991} Parker, E. N., The optical analogy for vector fields, {\it Phys. Fluids} B 3, 2652, 1991.
\bibitem[]{Arnold1997}Arnold, V. I., K. Vogtmann \& A. Weinstein, {\it Mathematical Methods of Classical Mechanics}, (Springer-Verlag) 1997.
\bibitem[]{saffman1992} Saffman, P. G., {\it Vortex Dynamics} (Cambridge University Press), 1992.
\bibitem[]{moore1979} Moore, D. W., The spontaneous appearance of a singularity in the shape of an evolving vortex sheet, {\it Proc. Roy. Soc.}, 365, 105, 1979.
\bibitem[]{calflischorellana1986} Calflisch, R. E., \& O. F. Orellana, Long time existence of a slightly perturbed vortex sheet, {\it Comm. Pure Appl. Math}, 39, 807,1986.
\bibitem[]{arnold1965} Arnol’d, V., Sur la topologie des ecoulements stationaires des fluids parfaits, {\it C. R. Acad. Sci. Paris}, 261, 17, 1965.
\bibitem[]{moffatt1985} Moffatt, H. K, Magnetostatic equilibria and analogous Euler flows of arbitrarily complex topology. I - Fundamentals, {\it J Fluid Mech.}, 159, 359, 1985.
\bibitem[]{McWilliamsPlierl1979} McWilliams, J. W., \& G. R. Plierl, On the evolution of nonlinear isolated vortices, {\it J. Phys. Ocean.} 9, 1155, 1979.
\bibitem[]{evs2019} Engvold, O., Vial, J.-C., \& Skumanich, A., edits., {\it The Sun as a Guide to Stellar Physics}  (Elsevier), 2019.
\bibitem[]{CaligariMoreno-InsertisSchussler1995} Caligari, P., F. Moreno-Insertis \& M. Schussler, Emerging flux tubes in the solar convection zone. 1: Asymmetry, tilt, and emergence latitude, {\it ApJ} 441, 886, 1995.
\bibitem[]{yfan2009} Fan, Y., Magnetic fields in the solar convection zone, {\it Living Revs in Solar Phys.}, 6, 96, 2009.
\bibitem[]{HesterZhangDikpati2025} Hester, R., J. Zhang \& M. Dikpati, A dynamic equilibrium theory for zonal circulation in the solar convection zone, {\it MNRAS} 538, 165, 2025.
\bibitem[]{ChoudhuriSchusslerDikpati1995} Choudhuri, A. R., M. Schussler \& M. Dikpati, The solar dynamo with meridional circulation, {\it Astron. Astrophys.} 303, L29, 1995.
\bibitem[]{DikpatiCharbonneau1999} Dikpati, M., \& P. Charbonneau, A Babcock-Leighton flux transport dynamo with aolar-like differential rotation, {\it ApJ} 518, 508, 1999.
\bibitem[]{yfanfang2014} Fan, Y.,\& F. Fang, A simulation of convective dynamo in the solar convective envelope: maintenance of the solar-like differential rotation and emerging flux, {\it ApJ} 789, 13, 2014.
\bibitem[]{McIntoshLeamon2024} McIntosh, S. W., \& R. J. Leaman, Deciphering solar magnetic activity: some (unpopular) thoughts on the coupling of the Sun's "weather" and "climate”, {\it Frontiers Astron. and Sp. Scien.}, 11, id. 1440708, 2024.
\bibitem[]{McIntoshLeamonEgeland2019} McIntosh, S. W., R. J. Leaman \& R. Egeland, What the sudden death of solar cycles can tell us about the nature of the solar interior, {\it Solar Phys.}, 294, 24, 2019.
\bibitem[]{Hathaway2015} Hathaway, D. H.,The solar cycle, {\it Living Rev. Solar Phys.} 12, 4, 2015.
\bibitem[]{Lites1995} Lites, B. W., B. C. Low, V. Martinez, et al., The possible ascent of a closed magnetic system through the photosphere, {\it ApJ} 446, 877, 1995.
\bibitem[]{gibsonfanmandrini2004} Gibson, S. E., Y. Fan \& C. Mandrini, Observational consequences of a magnetic flux rope emerging into the corona, {\it ApJ} 617, 600, 2004.
\bibitem[]{ZhangFlyerLow2006} Zhang, M., N. Flyer, \& B. C. Low, Magnetic field confinement in the corona: The role of magnetic helicity accumulation, {\it ApJ} 644, 575, 2006.
\bibitem[]{Raouafi2009} Raouafi, N. E., Observational evidence for coronal twisted flux rope, {\it ApJL}, 691, L28, 2009.
\bibitem[]{Berger2011} Berger, T., P. Testa, A. Hillier, et al., Magneto-thermal convection in solar prominences, {\it Nature} 472, 197, 2011.
\bibitem[]{Chen2025} Chen, P. F., Solar filament physiognomy: inferring magnetic quantities from imaging observations, {\it Solar Phys.} 300, 172, 2025.
\bibitem[]{gibsonLow1998} Gibson, S. E., \& B. C. Low, A time-dependent, three dimensional magnetohydrodynamic model of the coronal mass ejection, {\it ApJ} 493, 460, 1998. 
\bibitem[]{fangibson2004} Fan, Y., \& S. E. Gibson, Numerical simulations of three-dimensional coronal magnetic fields resulting from the emergence of twisted magnetic flux tubes, {\it ApJ} 609, 1123, 2004.
\bibitem[]{manchest2004} Manchester IV, W., T. Gombosi \& F. Y. DeZeeuw, Eruption of a buoyant emergent magnetic flux rope, {\it ApJ} 610, 588, 2004.
\bibitem[]{zhanglow2005} Zhang, M., \& B. C. Low, The hydromagnetic nature of solar Coronal Mass Ejections, {\it Ann. Rev. Astron. Astrophys.} 43, 103, 2005.
\bibitem[]{fangibson2007} Fan, Y., \& S. E. Gibson, Onset of Coronal Mass Ejections due to loss of confinement of coronal flux ropes, {\it ApJ} 668, 1232, 2007.
\bibitem[]{parker1963} Parker, E. N., Interplanetary Dynamical Process, {\it Interscience}, 1963.
\bibitem[]{low1997} Low, B. C., The role of Coronal Mass Ejections in solar activity, {\it Geophys. Monogr.}, 99, 39, 1997.
\bibitem[]{Gopal2003} Gopalswamy, N., A. Lara, S. Yashiro \& R. A. Howard, Coronal mass ejections and solar polarity reversal, {\it ApJ} 598, L63, 2003.
\bibitem[]{Low1994} Low, B. C., Magnetohydrodynamic processes in the solar corona: Flares, coronal mass ejections and magnetic helicity, {\it Phys. Plasama} 1,1684, 1994.
\bibitem[]{Rust1994} Rust, D. M., Spawning and shedding helical fields in the solar atmosphere, {Geophys. Res. Lett.} 21, 241, 1994.
\bibitem[]{DStLowthesis} De Sterck, H., B. C. Low \& S. Poedts, Complex magnetohydrodynamic bow shock topology in field-aligned low-$\beta$ flow around a perfectly conducting cylinder, {\it Phys. Plasmas} 5, 4015, 1998.
\bibitem[]{desterckpodts1999} De Sterck, H., \& S. Poedts, Field-aligned magnetohydrodynamic bow shock flows in the switch-on regime, {\it Astron. Astrophys.} 343, 641,1999.
\bibitem[]{KerrBrandenburg1999} Kerr, R. M., \& A. Brandenburg, Evidence for a singularity in ideal magnetohydrodynamics: Implications for fast reconnection. {\it Phys. Rev.Lett.} 83,1155, 1999.
\bibitem[]{GrafkeHomannDreher2008} Grafke, T., H. Homann, et. al. Numerical simulations of possible finite time singularities in the incompressible Euler equations: Comparison of numerical methods, {\it Physica D.}237, 1932, 2008.
\bibitem[]{ramitlowsmolar2010}Bhattacharyya, R., B. C. Low, \& P. K. Smolarkiewicz, On spontaneous formation of current sheets: Untwisted magnetic fields, {\it Phys. Plasmas} 17, 112901, 2010.
\bibitem[]{stone1999}  Stone, J. M., The ZEUS code for astrophysical magnetohydrodynamics: new extensions and applications, {\it J. Comp. Applied Math.} 109, 261, 1999.


%%%%%%%%%%%%%%%%%%%%%%%%%%%%%%%%%%%%%%%%%
\end{thebibliography}

\end{document}